\documentclass[11pt,a4paper]{article}
\pdfoutput=1
\usepackage{jheppub}
\usepackage{physics}
\usepackage{empheq}
\usepackage{longtable}
\usepackage{afterpage}

\usepackage{slashed}
\usepackage[normalem]{ulem}

\newcommand\nn{\nonumber}
\newcommand\fft[2]{\frac{#1}{#2}}
\usepackage{xcolor}

\preprint{LCTP-21-02}

\title{The Bethe-Ansatz approach to the $\mathcal N=4$ superconformal index at finite rank}

\author[a,b,c]{\small{Alfredo Gonz\'alez Lezcano}}
\emailAdd{agonzale@ictp.it}
\affiliation[a]{The Abdus Salam International Centre for Theoretical Physics, 34014 Trieste, Italy}
\affiliation[b]{SISSA International School for Advanced Studies\\
Via Bonomea 265, 34136 Trieste\\
\hspace*{15mm}  and\\
INFN, sezione di Trieste}
\affiliation[c]{Departamento de F\'isica, Universidad de Pinar del R\'io
Avenida Jos\'e Mart\'i No. 270,\\ CP 20100, Pinar del R\'io, Cuba }

\author[d]{\small{Junho Hong}}
\emailAdd{junhoh@umich.edu}

\author[d]{\small{James T. Liu}}
\emailAdd{jimliu@umich.edu}

\affiliation[d]{Leinweber Center for Theoretical Physics, University of Michigan, Ann Arbor, MI 48109, U.S.A.}

\author[a, d]{\small{Leopoldo A. Pando Zayas}}
\emailAdd{lpandoz@umich.edu}

\abstract{We investigate the Bethe-Ansatz approach to the superconformal index of ${\cal N}=4$ supersymmetric Yang-Mills with SU($N$) gauge group in the context of finite rank, $N$. We explicitly explore the role of the various types of solutions to the Bethe-Ansatz Equations in recovering the exact index for $N=2,3$. We classify the Bethe-Ansatz Equations solutions as standard (corresponding to a freely acting orbifold $T^2/\mathbb{Z}_m \times \mathbb{Z}_n$) and non-standard. For $N=2$, we find that the index is  fully recovered by  standard solutions and displays an interesting pattern of cancellations. However, for $N\ge 3$, the standard solutions alone do not suffice to reconstruct the index. We present quantitative arguments in various regimes of fugacities that highlight the challenging role played by the continuous families of non-standard solutions.}

\arxivnumber{}

\begin{document}

\maketitle

\newpage
\section{Introduction}
An insightful analysis of the superconformal index of ${\cal N}=4$ maximally supersymmetric Yang-Mills theory with gauge group SU($N$) has recently provided a microscopic foundation for the entropy of electrically charged, rotating, asymptotically AdS$_5$ black holes \cite{Cabo-Bizet:2018ehj, Choi:2018hmj, Benini:2018ywd}. The results are an important improvement on the understanding of the superconformal index previously introduced in \cite{Romelsberger:2005eg,Kinney:2005ej} and provide an explicit realization of a conjecture put forward in \cite{Hosseini:2017mds} regarding the entropy of AdS$_5$ black holes. These developments motivated various studies into the superconformal index of large classes of 4d ${\cal N}=1$ theories  \cite{Honda:2019cio,ArabiArdehali:2019tdm,Kim:2019yrz,Cabo-Bizet:2019osg,Amariti:2019mgp,Lezcano:2019pae,Lanir:2019abx,ArabiArdehali:2019orz}.

The problem of microscopic counting of the entropy has thus descended into a technical plane. Two main technical approaches have emerged, one rooted in saddle point approximations \cite{Cabo-Bizet:2018ehj, Choi:2018hmj}, and one in a Bethe-Ansatz (BA) formula of the index \cite{Benini:2018ywd}; a systematic discussion comparing both approaches including sub-leading contributions and extending the results to include 4d ${\cal N}=1$ theories was presented in \cite{GonzalezLezcano:2020yeb}. Other approaches to the evaluation of the index, include, for example, those rooted in doubly-periodic extensions \cite{Cabo-Bizet:2020nkr,Cabo-Bizet:2019eaf}, direct numerical evaluation \cite{Murthy:2020rbd,Agarwal:2020zwm}; a partial list of results includes \cite{Copetti:2020dil,Goldstein:2020yvj,Cabo-Bizet:2020ewf,Amariti:2020jyx}.

In the context of the AdS/CFT correspondence the superconformal index (SCI) represents the full quantum entropy of the dual black holes. The drive to have an exact in $N$ expression for the index is motivated by nothing less than to have the exact quantum entropy of the dual black holes. One expects such object will have powerful lessons to teach us about the nature of quantum gravity.  Indeed, studies of the superconformal index have already yielded important insight into aspects of quantum gravity in asymptotically AdS spacetimes. The analysis of \cite{GonzalezLezcano:2020yeb} and more recently \cite{Amariti:2020jyx}, have established the robustness of the ultraviolet prediction for the logarithmic corrections to the entropy; a study reported in \cite{Aharony:talk} has provided some insight into the structure of certain non-perturbative terms in the index. More ambitiously, one can expect that the BA structure of the index, whereby it is rewritten as the contributions of solutions to the Bethe-Ansatz Equations (BAE), might give us some clues about the path integral on the gravity side. Similarly, in the saddle point approach to the index one might expect that the hierarchy of saddle point solutions is related to the contributions to the gravitational path integral. With that motivation in mind, we explore the extent to which the exact superconformal index can be reconstructed from the BAE solutions. 
    
In this manuscript we explore, within the BA approach to the SCI, the ingredients necessary to reconstruct the full exact index. Most of the recent work devoted to the SCI has been analytic in nature.  There have been, however, two recent studies exploiting a direct numerical approach to the index with the main goal of better understanding finite $N$  aspects \cite{Murthy:2020rbd,Agarwal:2020zwm}. In part inspired by these developments we study the SCI at finite $N$ focusing on $N=2,3$. We take advantage of the relative simplicity of these cases to shed light on general aspects of the BA approach to recovering the full index. 
    
The BA approach to the superconformal index has the advantage of providing, in principle, an exact formula for the index. The devil, as we know, is in the details and the details in this context are the set of BAE solutions that need to be included to compute the index. Roughly speaking, following the classification of \cite{Hong:2018viz,ArabiArdehali:2019orz}, all the BAE solutions include standard (corresponding to a freely acting orbifold $T^2/\mathbb{Z}_m \times \mathbb{Z}_n$) and non-standard ones. One concrete result of this manuscript is to clarify the role that these two types of solutions play in reconstructing the full SCI.  The large $N$ picture is by now fairly clear, the dominant contribution coming from the so-called basic solutions and perhaps a small set of other solutions depending on the fugacities \cite{ArabiArdehali:2019orz,GonzalezLezcano:2020yeb}. In this manuscript we focus on the more subtle finite $N$ issues, in particular, we study how each type  contributes to the computation of the full SCI. In particular, we will demonstrate the important role played, for $N\ge 3$, by non-standard solutions. 

The rest of the manuscript is organized as follows. We start in Section \ref{sec:review} by briefly reviewing the BA approach to the superconformal index. We discuss the SU(2) case in Section \ref{sec:SU(2)} and SU(3) in Section \ref{sec:SU(3)}.  We discuss our results in Section \ref{sec:discussion} and relegate a number of technical details to a series of appendices.

\section{Review of the Bethe-Ansatz approach}\label{sec:review}
The superconformal index (SCI) of 4d $\mathcal N=4$ SU($N$) supersymmetric-Yang-Mills (SYM) theory \cite{Romelsberger:2005eg,Kinney:2005ej} is given as (following the convention of \cite{Benini:2018ywd}\footnote{We have used $Q_a^\text{here}=\fft12R_a^\text{there}~(a=1,2,3)$.}):
\begin{equation}
    \mathcal I(y_a,p,q) = {\rm Tr}(-1)^F e^{-\beta\{\mathcal{Q},\mathcal{Q}^\dagger\}} p^{J_1}q^{J_2}y_1^{Q_1}y_2^{Q_2}y_3^{Q_3}.\label{SCI:N=4}
\end{equation}
The SCI (\ref{SCI:N=4}) above receives contributions from the $\fft{1}{16}$-BPS states of the radially quantized theory on $\mathbb R\times S^3$ that preserve a complex supercharge $\mathcal Q$. These BPS states are characterized by the charges $J_{1,2}$ and $Q_{1,2,3}$. Here $J_{1,2}=J_L\pm J_R$ are angular momenta associated with SU(2)$_L\times$SO(2)$_R\cong$ SO(4) acting on $S^3$ and $Q_{1,2,3}$ are three $R$-charges for U(1)$^3\subset$ SO(6)$_R$. The fugacities $p,q,y_{1,2,3}$ are associated with the quantum numbers $J_{1,2}$ and $Q_{1,2,3}$ respectively, and constrained as $pq=y_1y_2y_3$. The SCI (\ref{SCI:N=4}) is well defined for $|p|,|q|<1$. 

One can rewrite the expression (\ref{SCI:N=4}) more explicitly in terms of elliptic hypergeometric integrals as \cite{Dolan:2008qi,Spiridonov:2010qv}  
\begin{equation}
    \mathcal I(y_a,p,q)=\fft{((p;p)_\infty(q;q)_\infty)^{N-1}}{N!}\prod_{a=1}^3\Gamma(y_a;p,q)^{N-1}\oint\prod_{i=1}^{N-1}\fft{dz_i}{2\pi iz_i}\prod_{i,j=1\,(i\neq j)}^N\fft{\prod_{a=1}^3\Gamma(\fft{z_i}{z_j}y_a;p,q)}{\Gamma(\fft{z_i}{z_j};p,q)},\label{SCI:N=4:integral}
\end{equation}
where the $z_i$-integration is over a unit circle with the SU($N$) constraint $\prod_{i=1}^Nz_i=1$. The integral (\ref{SCI:N=4:integral}) can be evaluated using a saddle point approximation \cite{Cabo-Bizet:2018ehj,Choi:2018hmj} (see also \cite{Honda:2019cio,ArabiArdehali:2019tdm,ArabiArdehali:2019orz}). It has recently been computed beyond the saddle point approximation and shown to be given, up to exponentially suppressed contributions, in terms of the exact $S^3$ partition function of a Chern-Simons theory \cite{GonzalezLezcano:2020yeb} (see also  \cite{Amariti:2020jyx} for more general classical groups).

One can also compute the SCI (\ref{SCI:N=4:integral}) following the Bethe-Ansatz (BA) approach. The BA approach has been introduced for a generic  4d $\mathcal N=1$ supersymmetric gauge theories in \cite{Benini:2018mlo}  based on insightful observations made in \cite{Closset:2017bse,Closset:2017zgf}. It was then applied to the $\mathcal N=4$ SU($N$) SYM theory with $p=q$ in \cite{Benini:2018ywd} and, more recently, to the same theory with $p=h^a$ and $q=h^b$ where $a,b\in\mathbb N$ and $\gcd(a,b)=1$ \cite{Benini:2020gjh}. Some discussion of the BA approach to the SCI of a large class of 4d ${\cal N}=1$ supersymmetric quiver gauge theories was presented to leading order in \cite{Lezcano:2019pae,Lanir:2019abx} and a systematic sub-leading study was presented recently in \cite{GonzalezLezcano:2020yeb}. In this manuscript we focus on the BA approach to the index of ${\cal N}=4$ SU($N$) SYM  with emphasis on aspects of the finite rank, $N$.   Our starting point is the corresponding presentation of the index following the BA formula for the SCI of the $\mathcal N=4$ SU($N$) SYM theory as (see Appendix \ref{App:functions} for the definitions of elliptic functions)
\begin{equation}
    \mathcal I(y_a,p,q)=\kappa(y_a,p,q)\sum_{\{u_i\}\in\mathcal M_\text{BAE}}\mathcal Z_{\text{tot}}(\{u_i\};\Delta,\sigma,\tau)H(\{u_i\};\Delta,\omega)^{-1},\label{eq:SCI:BA}
\end{equation}
where $z_i=e^{2\pi i u_i}$, $y_a=e^{2\pi i\Delta_a}$, $p=h^a=e^{2\pi i\sigma}$, $q=h^b=e^{2\pi i\tau}$, $h=e^{2\pi i\omega}$, and 
\begin{subequations}
\begin{align}
    \kappa(y_a,p,q)&=\fft1{N!}\left((p; p)_\infty(q;q)_\infty\prod_{a=1}^3\Gamma(y_a,p,q)\right)^{N-1},\\
    \mathcal Z_{\text{tot}}(\{u_i\};\Delta,\sigma,\tau) & =\sum_{m_1=1}^{ab}\cdots\sum_{m_{N-1}=1}^{ab}\mathcal{Z}(\{u_l-m_l\omega\};\Delta,\sigma,\tau)\label{eq:Ztot}\\
    \mathcal{Z}(\{u_i\};\Delta,\sigma,\tau) &=\prod_{i\ne j}^N\left(\widetilde\Gamma(u_{ij};\sigma,\tau)^{-1}\prod_{a=1}^3\widetilde\Gamma(u_{ij}+\Delta_a;\sigma,\tau)\right),\label{eq:Z}\\
    H(\{u_i\};\Delta,\omega)&=\det\left[\fft1{2\pi i}\fft{\partial(Q_1,\ldots, Q_N)}{\partial(u_1,\ldots,u_{N-1},\lambda)}\right].\label{eq:H}
\end{align}\label{building:blocks}%
\end{subequations}
Here $u_{ij}\equiv u_i-u_j$ and $\{u_i\}$ is a shorthand notation for $N$ holonomies $\{u_i|\,i=1,\cdots,N\}$. The SU($N$) constraint is given as $\sum_{i=1}^Nu_i\in\mathbb Z$. In (\ref{eq:Ztot}), the $N$-th integer $m_N$ is determined through the constraint
\begin{align}
    \sum_{l=1}^{N}m_l=0.
\end{align}
The BA operator used in (\ref{eq:H}) is defined as
\begin{equation}
\begin{split}
    Q_i(\{u_j\};\Delta,\omega)&\equiv e^{2\pi i(\lambda+3\sum_{j=1}^Nu_{ij})}\prod_\Delta\prod_{j=1}^N\fft{\theta_0(u_{ji}+\Delta;\omega)}{\theta_0(u_{ij}+\Delta;\omega)}\\
    &=e^{2\pi i\lambda}\prod_\Delta\prod_{j=1}^N\fft{\theta_1(u_{ji}+\Delta;\omega)}{\theta_1(u_{ij}+\Delta;\omega)},\label{eq:Q}
\end{split}
\end{equation}
where $\Delta$ take values in $\Delta\in\{\Delta_1,\Delta_2,-\Delta_1-\Delta_2\}$ and $\lambda$ is a free parameter that will be determined next. The Bethe-Ansatz Equations (BAE) used in the BA formula (\ref{eq:SCI:BA}) is then given as a system of transcendental equations as
\begin{equation}
    Q_i(\{u_j\};\Delta,\omega)=1,\label{eq:BAE}
\end{equation}
which fixes the parameter $e^{2\pi i\lambda}$ to a $N$-th root of unity. $\mathcal M_\text{BAE}$ in (\ref{eq:SCI:BA}) denotes a set of BAE solutions whose first $N-1$ holonomies are within the fundamental domain, namely
\begin{equation}
    \{u_i\}\in\mathcal M_\text{BAE}\quad\text{iff}\quad
    \begin{aligned}
    &i)~Q_i(\{u_j\};\Delta,\omega)=1\quad(i=1,\cdots,N)\\
    &ii)~u_i=x_i+y_i\omega~~\text{with}~~0\leq y_i<1\quad(i=1,\cdots,N-1)
    \end{aligned}~.\label{M:BAE}
\end{equation}

For later purpose, here we summarize the key properties of the BA operator (\ref{eq:Q}) and the building blocks (\ref{building:blocks}). Let us begin with the  BA operator (\ref{eq:Q}) which is doubly periodic with respect to holonomies and invariant under a constant shift as follows
\begin{subequations}
\begin{align}
    Q_i(\{u_j+m_j+n_j\omega\};\Delta,\omega)&=Q_i(\{u_j\};\Delta,\omega)\qquad(n_j,m_j\in\mathbb Z),\label{Q:double:period}\\
    Q_i(\{u_j+c\};\Delta,\omega)&=Q_i(\{u_j\};\Delta,\omega).\label{Q:constant:shift}
\end{align}\label{Q:properties}%
\end{subequations}
One of the building blocks $\mathcal Z(\{u_i\};\Delta,a\omega,b\omega)$ (\ref{eq:Z}) in the BA formula (\ref{eq:SCI:BA}) is quasi-periodic with respect to holonomies and invariant under a constant shift as
\begin{subequations}
\begin{align}
    \mathcal Z(\{u_j-\delta_{jk}ab\omega\};\Delta,\sigma,\tau)&=(-1)^{N-1}e^{-2\pi i\lambda}Q_k(\{u_j\};\Delta,\omega)\mathcal Z(\{u_j\};\Delta,\sigma,\tau),\label{Z:quasi:period}\\
    \mathcal Z(\{u_j+c\};\Delta,\sigma,\tau)&=\mathcal Z(\{u_j\};\Delta,\sigma,\tau).\label{Z:constant:shift}
\end{align}\label{Z:properties}%
\end{subequations}
The quasi-periodicity can be proved using (\ref{theta0:periodic}), (\ref{theta0:inversion}), and (\ref{Gamma:periodic:2}). Another building block $H(\{u_i\};\Delta,\omega)$ (\ref{eq:H}) in the BA formula (\ref{eq:SCI:BA}) satisfies similar properties explicitly given as:
\begin{subequations}
\begin{align}
    H(\{u_j-\delta_{jk}\omega\};\Delta,\sigma,\tau)&=H(\{u_j\};\Delta,\sigma,\tau),\label{H:period}\\
    H(\{u_j+c\};\Delta,\sigma,\tau)&=H(\{u_j\};\Delta,\sigma,\tau),\label{H:constant:shift}
\end{align}\label{H:properties}%
\end{subequations}
which follows from its definition in (\ref{eq:H}) and the properties (\ref{Q:properties}). Note that, according to (\ref{H:period}), the determinant $H(\{u_i\};\Delta,\omega)$ is invariant under the $\omega$-shift of the $k$-th holonomy with an arbitrary $k=1,\cdots,N$. Hence it is periodic with respect to holonomies, which is distinguished from the quasi-periodicity of a building block $\mathcal Z(\{u_i\};\Delta,a\omega,b\omega)$ given in (\ref{Z:quasi:period}). For SU($N$), the determinant (\ref{eq:H}) reduces to that of an $(N-1)\times(N-1)$ matrix $\mathbb H$ \cite{Hong:2018viz,GonzalezLezcano:2020yeb}
\begin{equation}
    H(\{u_i\};\Delta,\omega)=N\det[\mathbb H_{ij}]\equiv N\det\left[\fft1{2\pi i}\fft{\partial(Q_1,\ldots, Q_{N-1})}{\partial(u_1,\ldots,u_{N-1})}\right].\label{eq:H:SU(N)}
\end{equation}
Evaluating $\partial Q_i/\partial u_j$ with the BA operator (\ref{eq:Q}) at BAE solutions satisfying (\ref{eq:BAE}), we obtain the elements of the matrix $\mathbb H$ explicitly as
\begin{equation}
\begin{split}
    \mathbb H_{ii}&=-\sum_{k\ne i}^{N-1}g(u_{ki};\omega)-2g(u_{Ni};\omega),\\
    \mathbb H_{ij}&=g(u_{ij};\omega)-g(u_{Nj};\omega)\qquad(i\ne j),
\end{split}
\end{equation}
where
\begin{equation}
    g(u;\omega)\equiv\fft{1}{2\pi i}\sum_\Delta\fft\partial{\partial\Delta}\left[\log\theta_1(u+\Delta;\omega)+\log\theta_1(-u+\Delta;\omega)\right].
\end{equation}
%

\section{The SU(2) index}\label{sec:SU(2)}
In this section, we specialize to the case of SU(2) with the goal of achieving a clear picture of how the full SCI arises  from the BA formula (\ref{eq:SCI:BA}) and the explicit role of the Bethe vacua (\ref{M:BAE}). We will further directly compare the results of the BA approach with other approaches such as the series expansion by counting states \cite{Murthy:2020rbd,Agarwal:2020zwm} and direct numerical integration of (\ref{SCI:N=4:integral}). 

For $N=2$, the BAE (\ref{eq:BAE}) reduces to a single transcendental equation as
\begin{equation}
    \pm1=e^{-2\pi i\lambda}=\prod_\Delta\fft{\theta_1(\Delta+u_{21};\omega)}{\theta_1(\Delta-u_{21};\omega)}\quad\Leftrightarrow\quad 1=\prod_\Delta\fft{\theta_1(\Delta+u_{21};\omega)^2}{\theta_1(\Delta-u_{21};\omega)^2},\label{eq:BAE:N=2}
\end{equation}
where $\Delta$ take values in $\Delta\in\{\Delta_1,\Delta_2,-\Delta_1-\Delta_2\}$. Note that the double-periodicity of the BA operator (\ref{Q:double:period}) implies that given a solution $u_{21}$, we can generate countably many BAE solutions $u_{21}+m+n\omega~(m,n\in\mathbb Z)$. For the SU(2) case at hand, if we identify solutions in different lattices as $u_{21}\sim u_{21}+\mathbb Z+\mathbb Z\omega$,  there are only 6 distinct BAE solutions \cite{ArabiArdehali:2019orz}. They are given as
\begin{equation}
    u_{21}\in\left\{0,\fft12,\fft\omega2,\fft{1+\omega}{2},u_\Delta,-u_\Delta\right\}.\label{sol:BAE:N=2}
\end{equation}
Let us now introduce a classification of the solutions. We will roughly group the solutoins to the BAE as standard and non-standard as follows (except the trivial one):
\begin{itemize}
    \item Standard solutions: the BAE solutions that correspond to a freely acting orbifold $T^2/\mathbb{Z}_m \times \mathbb{Z}_n$.These solutions can be associated to an $SL(2,\mathbb{Z})$ action and are  generically  $\Delta$-indepent. 
    \item Non-standard solutions: All the other solutions, they have the generic property of being $\Delta$-dependent.
\end{itemize}
Applied to the set of solutions in (\ref{sol:BAE:N=2}), the first solution is the trivial one and the next three are \underline{standard} solutions denoted by three integers as $\{2,1,0\}$, $\{1,2,0\}$, and $\{1,2,1\}$ respectively in the convention of \cite{Hong:2018viz}. The last two are called \underline{non-standard} solutions that depend on chemical potentials \cite{ArabiArdehali:2019orz}. The explicit form of a non-standard solution is known only for real $\Delta_{1,2}$ in the asymptotic regions: for example, in the `low-temperature' limit $|\omega|\to\infty$ with fixed $\arg\omega$, $u_\Delta$ is given for real $\Delta_{1,2}$ as \cite{ArabiArdehali:2019orz}
\begin{equation}
    u_\Delta(|\omega|\to\infty)=\fft{1}{2\pi i}\log\left[-\left(\fft{1-\sum_\Delta\cos2\pi\Delta}{2}\right)+\sqrt{\left(\fft{1-\sum_\Delta\cos2\pi\Delta}{2}\right)^2-1}\right].\label{NS:low-temp:N=2}
\end{equation}
It is also convenient to characterize the solutions based on the value of 
 $e^{-2\pi i\lambda}$ in equation (\ref{eq:BAE:N=2}). In this case we  can split the 6 solutions (\ref{sol:BAE:N=2}) into two groups as
\begin{subequations}
\begin{align}
    e^{-2\pi i\lambda}=-1\quad&:\quad u_{21}\in\left\{\fft12,\fft\omega2,\fft{1+\omega}{2}\right\},\label{sol:BAE:N=2:yes}\\
    e^{-2\pi i\lambda}=1\quad&:\quad u_{21}\in\left\{0,u_\Delta,-u_\Delta\right\}.\label{sol:BAE:N=2:no}
\end{align}\label{sol:BAE:N=2:groups}%
\end{subequations}
It is noteworthy that, in the SU(2) case, the characterization by the value of $e^{-2\pi i\lambda}$ (\ref{sol:BAE:N=2:groups}) distinguishes standard solutions (\ref{sol:BAE:N=2:yes}) from a trivial one and non-standard ones (\ref{sol:BAE:N=2:no}). We will see that such grouping will play an important role in computing the SU(2) index through the BA formula (\ref{eq:SCI:BA}) below. 

Observe that the low-temperature asymptotic form (\ref{NS:low-temp:N=2}) is enough to evaluate the value of $e^{-2\pi i\lambda}$ for a non-standard solution $u_{21}=\pm u_\Delta$. This is because the value of $e^{-2\pi i\lambda}$ cannot jump between $\pm1$ under continuous deformation of $\omega$: once the value of $e^{-2\pi i\lambda}$ is determined in the low-temperature limit $|\omega|\to\infty$, it has to be the same for arbitrary $\omega$.

Now we consider the contribution from a BAE solution $u_{21}=u^\star$, which is an arbitrary element of the 6 solutions listed in (\ref{sol:BAE:N=2}), to the SCI through the BA formula (\ref{eq:SCI:BA}). Using the double-periodicity of the BA operator (\ref{Q:double:period}), we set 
\begin{equation}
    u^\star=x^\star+y^\star\omega\qquad\text{with}\quad-1<y^\star\leq0
\end{equation}
without loss of generality. From this BAE solution $u_{21}=u^\star$, we can generate a total of 4 inequivalent elements $\{u_1,u_2\}$ of $\mathcal M_\text{BAE}$ (\ref{M:BAE}) using the properties of the BA operator (\ref{Q:properties}) as
\begin{equation}
    \mathcal M_\text{BAE}\ni\{-\fft{u^\star}{2}+\fft{r+s_1\omega}{2},\fft{u^\star}{2}+\fft{r+s_2\omega}{2}\},
\end{equation}
where
\begin{equation}
    r\in\{0,1\},\qquad \{s_1,s_2\}\in\{\{0,0\},\{1,-1\}\}.
\end{equation}
Substituting these 4 elements into the BA formula (\ref{eq:SCI:BA}) and using (\ref{Z:constant:shift}), we obtain the contribution from a BAE solution $u_{21}=u^\star$ to the SCI
\begin{equation}
    \mathcal I_{\{u_{21}=u^\star\}}(y_a,p,q)=2\kappa(y_a,p,q)\sum_{m_1=1}^{2ab}\fft{\mathcal Z(\{-\fft{u^\star+m_1\omega}{2},\fft{u^\star+m_1\omega}{2}\};\Delta,a\omega,b\omega)}{H(\{-\fft{u^\star+m_1\omega}{2},\fft{u^\star+m_1\omega}{2}\};\Delta,\omega)}.\label{eq:SCI:u^star:N=2}
\end{equation}
Using the properties of $\mathcal Z(\{u_i\};\Delta,a\omega,b\omega)$ (\ref{Z:properties}) and $H(\{u_i\};\Delta,\omega)$ (\ref{H:properties}), we can simplify (\ref{eq:SCI:u^star:N=2}) further as
\begin{equation}
    \mathcal I_{\{u_{21}=u^\star\}}(y_a,p,q)=\begin{cases}
    4\kappa(y_a,p,q)\fft{\sum_{m_1=1}^{ab}\mathcal Z(\{-\fft{u^\star+m_1\omega}{2},\fft{u^\star+m_1\omega}{2}\};\Delta,a\omega,b\omega)}{H(\{-\fft{u^\star}{2},\fft{u^\star}{2}\};\Delta,\omega)} & (e^{-2\pi i\lambda}=-1)\\
    0 & (e^{-2\pi i\lambda}=1)
    \end{cases},\label{eq:SCI:u^star:N=2:simple}
\end{equation}
where the value of $e^{-2\pi i\lambda}$ follows from (\ref{sol:BAE:N=2:groups}) for a given $u_{21}=u^\star$.

Finally, the SU(2) index is given as the sum of (\ref{eq:SCI:u^star:N=2:simple}) over all BAE solutions $u_{21}=u^\star$ listed in (\ref{sol:BAE:N=2}). The result can be written strictly in terms of the standard solutions as
\begin{empheq}[box=\fbox]{equation}
    \mathcal I(y_a,p,q)=4\left(\mathcal I_{\{2,1,0\}}+\mathcal I_{\{1,2,0\}}+\mathcal I_{\{1,2,1\}}\right),\label{eq:SCI:BA:N=2}
\end{empheq}
where we have defined
\begin{subequations}
\begin{align}
    \kappa^{-1}\mathcal I_{\{2,1,0\}}&=\fft{\sum_{m_1=1}^{ab}\mathcal Z(\{-\fft{1/2+m_1\omega}{2},\fft{1/2+m_1\omega}{2}\};\Delta,a\omega,b\omega)}{H(\{-\fft14,\fft14\};\Delta,\omega)},\label{eq:SCI:BA:N=2:210}\\
    \kappa^{-1}\mathcal I_{\{1,2,0\}}&=\fft{\sum_{m_1=1}^{ab}\mathcal Z(\{-\fft{(m_1-1/2)\omega}{2},\fft{(m_1-1/2)\omega}{2}\};\Delta,a\omega,b\omega)}{H(\{-\fft\omega4,\fft\omega4\};\Delta,\omega)},\label{eq:SCI:BA:N=2:120}\\
    \kappa^{-1}\mathcal I_{\{1,2,1\}}&=\fft{\sum_{m_1=1}^{ab}\mathcal Z(\{-\fft{1/2+(m_1-1/2)\omega}{2},\fft{1/2+(m_1-1/2)\omega}{2}\};\Delta,a\omega,b\omega)}{H(\{-\fft{1+\omega}{4},\fft{1+\omega}{4}\};\Delta,\omega)}.\label{eq:SCI:BA:N=2:121}
\end{align}\label{eq:SCI:BA:N=2:standard}%
\end{subequations}
We remark that the non-standard solutions in the SU(2) case evaluate to zero but this is an accident of SU(2). The non-standard solutions will play a more prominent, albeit puzzling, role in generic cases of SU($N$).

\subsection{Asymptotic behavior}\label{sec:SU(2):asymp}
The SU(2) index (\ref{eq:SCI:BA:N=2}) is exact but each contribution from standard solutions listed in (\ref{eq:SCI:BA:N=2:standard}) is a complicated combination of elliptic Gamma functions.
Hence, in this subsection, we investigate the SU(2) index (\ref{eq:SCI:BA:N=2}) in the asymptotic regions where the exact BA formula (\ref{eq:SCI:BA:N=2}) can be written in terms of elementary functions. This will allow us to understand the behavior of the SU(2) index more intuitively and also compare it with results from other approaches including the series expansion by counting states \cite{Murthy:2020rbd,Agarwal:2020zwm} and a numerical integration of (\ref{SCI:N=4:integral}). The asymptotic regions we consider are  the low-temperature limit ($|\omega|\to\infty$ or $|h|\to0$) and the Cardy-like limit ($|\omega|\to0$ or $|h|\to1$) with fixed $\arg\omega$. 


\subsubsection{The low-temperature limit}\label{LowTSU2}
When $|h|<1$, we can expand the SU(2) index (\ref{eq:SCI:BA:N=2}) as a series in $h$, which gives a reasonable approximation under $|\omega|\to\infty$ or $|h|\to0$. Specializing to the case $p=q=h=y_a^{3/2}$, corresponding to $(a,b)=(1,1)$, and using the product representations of the Pochhammer symbol and the elliptic Gamma function, the contributions from the three standard solutions (\ref{eq:SCI:BA:N=2:standard}) can be expanded as
\begin{subequations}
\begin{align}
    \mathcal I_{\{2,1,0\}}&=-\frac{1}{2}-3  x-\frac{21}{2} x^2-31 x^3-87  x^4-225 
    x^5-\frac{1071}{2}x^6-1215  x^7-2661 x^8\nn\\&\quad-5598 
    x^9-\frac{22755}{2}x^{10}+\mathcal O\left(x^{11}\right),\label{eq:SCI:BA:N=2:210:lowT}\\
    \mathcal I_{\{1,2,0\}}&=-\frac{1}{16  x^{3/2}}-\frac{3}{16      x^{1/2}}+\frac{3}{8}-\frac{3}{4}
    x^{1/2}+\frac{3}{2}x-\frac{43}{16}
    x^{3/2}+\frac{21}{4}x^2-\frac{153}{16}  x^{5/2}\nn\\&\quad+\frac{31 
    }{2}x^3-\frac{105}{4} x^{7/2}+\frac{177}{4} x^4-\frac{1131}{16}
    x^{9/2}+\frac{447}{4} x^5-\frac{2775}{16} x^{11/2}+\frac{2135
    }{8} x^6\nn\\&\quad-\frac{1635}{4} x^{13/2}+\frac{2439}{4} x^7-\frac{7211}{8}
     x^{15/2}+\frac{5325}{4}  x^8-\frac{30999}{16}
    x^{17/2}+\frac{5589}{2} x^9\nn\\&\quad-\frac{16023}{4}  x^{19/2}+\frac{11379}{2}
     x^{10}-\frac{64327}{8} x^{21/2}+\mathcal O\left(x^{11}\right),\label{eq:SCI:BA:N=2:120:lowT}\\
    \mathcal I_{\{1,2,1\}}&=\frac{1}{16  x^{3/2}}+\frac{3}{16  x^{1/2}}+\frac{3}{8}+\frac{3}{4}
    x^{1/2}+\frac{3}{2}x+\frac{43}{16}
    x^{3/2}+\frac{21}{4} x^2+\frac{153}{16}  x^{5/2}\nn\\&\quad+\frac{31 }{2}x^3+\frac{105}{4} x^{7/2}+\frac{177}{4}  x^4+\frac{1131}{16}
    x^{9/2}+\frac{447}{4} x^5+\frac{2775}{16}  x^{11/2}+\frac{2135
     x^6}{8}\nn\\&\quad+\frac{1635}{4}x^{13/2}+\frac{2439}{4}  x^7+\frac{7211}{8}
     x^{15/2}+\frac{5325}{4} x^8+\frac{30999}{16} 
    x^{17/2}+\frac{5589}{2}x^9\nn\\&\quad+\frac{16023}{4}  x^{19/2}+\frac{11379}{2}
    x^{10}+\frac{64327}{8} x^{21/2}+\mathcal O\left(x^{11}\right).\label{eq:SCI:BA:N=2:121:lowT}
\end{align}\label{eq:SCI:BA:N=2:standard:lowT:1}%
\end{subequations}
Here we follow the notation of \cite{Murthy:2020rbd,Agarwal:2020zwm}, where $x$ is defined by the relations $p=q=x^3$ and $y_a=x^2$.  Combining these three standard solutions according to (\ref{eq:SCI:BA:N=2}) then gives the SU(2) index
\begin{equation}
    \mathcal I(y_a=x^2,p=x^3,q=x^3)=1+6x^4-6x^5-7x^6+18x^7+6x^8-36x^9+6x^{10}+\mathcal O(x^{11}),\label{I:low:identical}
\end{equation}
which agrees with the generalized series expansion of the SU(2) index (\ref{I:low:generic}) derived in Appendix \ref{App:lowT} based on \cite{Murthy:2020rbd,Agarwal:2020zwm}.  Although these expressions are presented here only up to $\mathcal O(x^{10})$, the Pochhammer symbol and elliptic Gamma function can easily be expanded to considerably higher order if desired.

We also investigated the more generic case $p=h^3$, $q=h^2$, and $y_a=h^{5/3}$ corresponding to $(a,b)=(3,2)$. In this case, the three standard solutions, (\ref{eq:SCI:BA:N=2:standard}), admit the expansions
\begin{subequations}
\begin{align}
    \mathcal I_{\{2,1,0\}}&=-\fft{3}{4x^3}-\fft{3}{x^2}-\fft{45}{4x}-33-\fft{381}{4}x-\fft{495}{2}x^2-\fft{2457}{4}x^3-1434x^4-\fft{12879}{4}x^5\nn\\
    &\quad-\fft{27747}{4}x^6-\fft{57867}{4}x^7-\fft{58509}{2}x^8-\fft{230523}{4}x^9-\fft{443271}{4}x^{10}-\fft{834345}{4}x^{11}\nn\\
    &\quad-\fft{1539861}{4}x^{12}+\cdots-\fft{36607515}{2}x^{19}+\mathcal O(x^{20}),\\
    \mathcal I_{\{1,2,0\}}&=-\fft{1}{8x^{9/2}}-\fft{3}{8x^{7/2}}+\fft{3}{8x^3}-\fft{9}{8x^{5/2}}+\fft{3}{2x^2}-\fft{13}{4x^{3/2}}+\fft{45}{8x}-\fft{39}{4x^{1/2}}+\fft{133}{8}\nn\\
    &\quad-\fft{57}{2}x^{1/2}+\fft{381}{8}x-77x^{3/2}+\cdots+\fft{73215039}{8}x^{19}-11782924x^{39/2}+\mathcal O(x^{20}),\\
    \mathcal I_{\{1,2,1\}}&=\fft{1}{8x^{9/2}}+\fft{3}{8x^{7/2}}+\fft{3}{8x^3}+\fft{9}{8x^{5/2}}+\fft{3}{2x^2}+\fft{13}{4x^{3/2}}+\fft{45}{8x}+\fft{39}{4x^{1/2}}+\fft{133}{8}\nn\\
    &\quad+\fft{57}{2}x^{1/2}+\fft{381}{8}x+77x^{3/2}+\cdots+\fft{73215039}{8}x^{19}+11782924x^{39/2}+\mathcal O(x^{20}),
\end{align}\label{eq:SCI:BA:N=2:standard:lowT:2}%
\end{subequations}
where we have taken $p=x^9$, $q=x^6$ and $y_a=x^5$.  Adding these contributions then gives the SU(2) index
\begin{equation}
    \mathcal I(y_a=x^5,p=x^9,q=x^6)=1+6x^{10}-3x^{11}-3x^{14}-7x^{15}+9x^{16}-3x^{17}+9x^{19}+\mathcal O(x^{20}).
\end{equation}
This agrees with the generalized series expansion of the SU(2) index, (\ref{I:low:generic}), when rearranged according to the scaling $p=h^3$, $q=h^2$, and $y_a=h^{5/3}$.

The above observation confirms that the BA formula (\ref{eq:SCI:BA}) is consistent with the series expansion (\ref{I:low:generic}) based on \cite{Murthy:2020rbd,Agarwal:2020zwm} in the low-temperature regime where $|h|<1$. This strongly supports that the $N=2$ BA formula (\ref{eq:SCI:BA:N=2}) gives the exact SU(2) index, in particular that the three contributions from standard solutions (\ref{eq:SCI:BA:N=2:standard}) are the only contributions to the SCI and each one has a degeneracy of 4. 

It is noteworthy that, both in (\ref{eq:SCI:BA:N=2:standard:lowT:1}) and (\ref{eq:SCI:BA:N=2:standard:lowT:2}), the series for $\mathcal I_{\{2,1,0\}}$ is a Taylor series in integer powers of $x$, but the other two series for $\mathcal I_{\{1,2,0\}}$ and $\mathcal I_{\{1,2,1\}}$ include half-integer powers of $x$. Moreover, they start at order $x^{-3/2}$. Remarkably, the half-integer powers of $x$ cancel in the sum $\mathcal I_{\{1,2,0\}}+\mathcal I_{\{1,2,1\}}$. From this observation in the SU(2) case, we expect $\mathcal I_{\{m,n,r\}}$, namely the contribution from a BAE solution to the SCI denoted by three integers $\{m,n,r\}$ following \cite{Hong:2018viz}, to be a series in powers of $x^{1/n}$ and that fractional powers of $x$ are removed in the sum $\sum_{r=0}^{n-1}\mathcal I_{\{m,n,r\}}$ in general. Nevertheless, inverse integer powers of $x$ can remain in this sum. If would be interesting to investigate whether this tantalizing  cancellations offer a bridge to the bootstrap ideas for the superconformal index based on modularity advanced by Gadde \cite{Gadde:2020bov}.

\subsubsection{The Cardy-like limit}\label{Cardy-SU2}
Next we investigate the Cardy-like limit ($|\omega|\to0$ or $|h|\to1$) of the SU(2) index through (\ref{eq:SCI:BA:N=2}). For simplicity, we identify $p=q$ with $(a,b)=(1,1)$. Standard contributions to the SU(2) index (\ref{eq:SCI:BA:N=2:standard}) are then written explicitly as
\begin{subequations}
\begin{align}
	\mathcal I_{\{2,1,0\}}&=\fft{(q;q)_\infty^2\prod_{a=1}^3\widetilde\Gamma(\Delta_a;\tau)}{\fft{4i}{\pi}\sum_\Delta\partial_\Delta\log[\theta_1(\fft12+\Delta;\tau)\theta_1(-\fft12+\Delta;\tau)]}\fft{\prod_{a=1}^3\widetilde\Gamma(\fft12+\Delta_a;\tau)\widetilde\Gamma(-\fft12+\Delta_a;\tau)}{\widetilde\Gamma(\fft12;\tau)\widetilde\Gamma(-\fft12;\tau)},\label{eq:SCI:BA:N=2:210:s=t}\\
	\mathcal I_{\{1,2,0\}}&=\fft{(q;q)_\infty^2\prod_{a=1}^3\widetilde\Gamma(\Delta_a;\tau)}{\fft{4i}{\pi}\sum_\Delta\partial_\Delta\log[\theta_1(\fft\tau2+\Delta;\tau)\theta_1(-\fft\tau2+\Delta;\tau)]}\fft{\prod_{a=1}^3\widetilde\Gamma(\fft\tau2+\Delta_a;\tau)\widetilde\Gamma(-\fft\tau2+\Delta_a;\tau)}{\widetilde\Gamma(\fft\tau2;\tau)\widetilde\Gamma(-\fft\tau2;\tau)},\label{eq:SCI:BA:N=2:120:s=t}\\
	\mathcal I_{\{1,2,1\}}&=\fft{(q;q)_\infty^2\prod_{a=1}^3\widetilde\Gamma(\Delta_a;\tau)}{\fft{4i}{\pi}\sum_\Delta\partial_\Delta\log[\theta_1(\fft{\tau+1}{2}+\Delta;\tau)\theta_1(-\fft{\tau+1}{2}+\Delta;\tau)]}\fft{\prod_{a=1}^3\widetilde\Gamma(\fft{\tau+1}{2}+\Delta_a;\tau)\widetilde\Gamma(-\fft{\tau+1}{2}+\Delta_a;\tau)}{\widetilde\Gamma(\fft{\tau+1}{2};\tau)\widetilde\Gamma(-\fft{\tau+1}{2};\tau)}.\label{eq:SCI:BA:N=2:121:s=t}
\end{align}\label{eq:SCI:BA:N=2:standard:s=t}%
\end{subequations}
From here on, we use $q$ and $\tau$ instead of $h$ and $\omega$ since they are the same under the identification $p=q=h$ with $(a,b)=(1,1)$. We will also use the ``$\sim$'' symbol for equations valid up to exponentially suppressed terms of the form $\mathcal O(e^{-1/|\tau|})$.

To begin with, substituting the asymptotic behaviors of $\theta_1(u;\tau)$ (\ref{elliptic:theta:1:asymp}) and $\widetilde\Gamma(u;\tau)$ (\ref{elliptic:Gamma:asymp}) into (\ref{eq:SCI:BA:N=2:120:s=t}) gives the Cardy-like limit of the contribution from the basic $\{1,2,0\}$ BAE solution as
\begin{equation}
    \log\mathcal I_{\{1,2,0\}}\sim-\fft{3\pi i}{\tau^2}\prod_{a=1}^3\left(\{\Delta_a\}_\tau-\fft{1+\eta_1}{2}\right)-\log2.\label{I:120:Cardy}
\end{equation}
Here the $\tau$-modded value $\{\cdot\}_\tau$ is defined in (\ref{tau-modded}) and we have introduced $\eta_C\in\{\pm1\}$ as
\begin{equation}
    \sum_{a=1}^3\{C\Delta_a\}_\tau=2C\tau+\fft{3+\eta_C}{2}\qquad\Leftrightarrow\qquad\sum_{a=1}^3\{C\tilde\Delta_a\}=\fft{3+\eta_C}{2},\label{eq:eta}
\end{equation}
assuming $\tilde\Delta_a\not\in\mathbb Z$. Refer to (\ref{u:component}) and (\ref{modded}) for the definitions of the `tilde' component of chemical potentials $\tilde\Delta_a$ and a real modded value $\{\cdot\}$ respectively.

For the other two BA contributions (\ref{eq:SCI:BA:N=2:210:s=t}) and (\ref{eq:SCI:BA:N=2:121:s=t}), we keep track of the leading exponentially suppressed terms since otherwise they diverge for the $\eta_1=\eta_2$ case. Substituting the asymptotic behaviors (\ref{elliptic:theta:1:asymp}) and (\ref{elliptic:Gamma:asymp}) into (\ref{eq:SCI:BA:N=2:210:s=t}) and (\ref{eq:SCI:BA:N=2:121:s=t}) then gives
\begin{equation}
\begin{split}
	&\log\mathcal I_{\{2,1,0\}}\\
	&\sim-\fft{\pi i}{2\tau^2}\prod_{a=1}^3\left(\{2\Delta_a\}_\tau-\fft{1+\eta_2}{2}\right)+\fft{\pi i}{\tau^2}\prod_{a=1}^3\left(\{\Delta_a\}_\tau-\fft{1+\eta_1}{2}\right)-\log 16\\
	&\quad+\sum_{a=1}^3\left(2\log\fft{\psi(\fft{\{1/2+\Delta_a\}_\tau}{\tau}-1)}{\psi(\fft{1-\{1/2+\Delta_a\}_\tau}{\tau}+1)}+\log\fft{\psi(\fft{\{\Delta_a\}_\tau}{\tau}-1)}{\psi(\fft{1-\{\Delta_a\}_\tau}{\tau}+1)}\right)+4\log(1-e^{-\fft{\pi i}{\tau}})\\
	&\quad+\begin{cases}
	\fft{3\eta_1\pi i}{4} & (\eta_1=-\eta_2)\\
	\fft{\pi i(6-5\eta_1)}{12}-\log\sum_\Delta\left(\fft{e^{-\fft{2\pi i}{\tau}(1-\{\fft12+\Delta\}_\tau)}}{1-e^{-\fft{2\pi i}{\tau}(1-\{\fft12+\Delta\}_\tau)}}-\fft{e^{-\fft{2\pi i}{\tau}\{\fft12+\Delta\}_\tau}}{1-e^{-\fft{2\pi i}{\tau}\{\fft12+\Delta\}_\tau}}\right) & (\eta_1=\eta_2)
	\end{cases},
\end{split}\label{I:210:Cardy}
\end{equation}
\begin{equation}
\begin{split}
	&\log\mathcal I_{\{1,2,1\}}\\
	&\sim-\fft{\pi i}{2\tau^2}\prod_{a=1}^3\left(\{2\Delta_a\}_\tau-\fft{1+\eta_2}{2}\right)+\fft{\pi i}{\tau^2}\prod_{a=1}^3\left(\{\Delta_a\}_\tau-\fft{1+\eta_1}{2}\right)-\log 16\\
	&\quad+\sum_{a=1}^3\left(\log\fft{\psi(\fft{\{1/2+\Delta_a\}_\tau}{\tau}-\fft12)}{\psi(\fft{1-\{1/2+\Delta_a\}_\tau}{\tau}+\fft12)}+\log\fft{\psi(\fft{\{1/2+\Delta_a\}_\tau}{\tau}-\fft32)}{\psi(\fft{1-\{1/2+\Delta_a\}_\tau}{\tau}+\fft32)}+\log\fft{\psi(\fft{\{\Delta_a\}_\tau}{\tau}-1)}{\psi(\fft{1-\{\Delta_a\}_\tau}{\tau}+1)}\right)\\
	&\quad+4\log(1+e^{\fft{\pi i}{\tau}})\\
	&\quad+\begin{cases}
	\fft{5\eta_1\pi i}{4} & (\eta_1=-\eta_2)\\
	\fft{\pi i(6-5\eta_1)}{12}-\log\sum_\Delta\left(\fft{-e^{-\fft{2\pi i}{\tau}(1-\{\fft12+\Delta\}_\tau)}}{1+e^{-\fft{2\pi i}{\tau}(1-\{\fft12+\Delta\}_\tau)}}-\fft{-e^{-\fft{2\pi i}{\tau}\{\fft12+\Delta\}_\tau}}{1+e^{-\fft{2\pi i}{\tau}\{\fft12+\Delta\}_\tau}}\right) & (\eta_1=\eta_2)
	\end{cases}.
\end{split}\label{I:121:Cardy}
\end{equation}
Refer to Appendix \ref{App:Cardy:N=2} for details. Note that the BA contributions (\ref{I:210:Cardy}) and (\ref{I:121:Cardy}) have the same $\fft{1}{\tau^2}$-leading order terms. The sub-leading terms are different, however, and this difference will play an important role in estimating the Cardy-like asymptotics of the SU(2) index in the region of chemical potentials dubbed as ``$W$-wing'' defined below in \eqref{Wings}. 

Now, substituting (\ref{I:120:Cardy}), (\ref{I:210:Cardy}), and (\ref{I:121:Cardy}) into (\ref{eq:SCI:BA:N=2}) gives the Cardy-like limit of the SU(2) index. Following the classification of \cite{ArabiArdehali:2019orz}, we investigate the resulting SU(2) index in the ``$M$-wing'' and in the ``$W$-wing,'' respectively . The $M$-wing is the region of chemical potentials where the contribution from the basic $\{1,N,0\}$ BAE solution, namely (\ref{I:120:Cardy}) for $N=2$, is dominant. In the $W$-wing the contribution from the basic solution is exponentially suppressed. These two regimes of chemical potentials are explicitly determined as 
\begin{equation}
\begin{split}
    M\text{-wing: }\Re\left[-\fft{i}{\tau^2}\prod_{a=1}^3\left(\{\Delta_a\}_\tau-\fft{1+\eta_1}{2}\right)\right]>0,\\
    W\text{-wing: }\Re\left[-\fft{i}{\tau^2}\prod_{a=1}^3\left(\{\Delta_a\}_\tau-\fft{1+\eta_1}{2}\right)\right]<0.
\end{split}\label{Wings}
\end{equation}
%

\subsubsection*{The Cardy-like limit in the $M$-wing}
In the $M$-wing, we can simplify (\ref{eq:SCI:BA:N=2}) with $p=q$ as
\begin{equation}
    \mathcal I(y_a,q,q)=4\mathcal I_{\{1,2,0\}}\left(1+\fft{\mathcal I_{\{2,1,0\}}+\mathcal I_{\{1,2,1\}}}{\mathcal I_{\{1,2,0\}}}\right)\sim 4\mathcal I_{\{1,2,0\}}.
\end{equation}
The SU(2) index is then given from (\ref{I:120:Cardy}) as
\begin{empheq}[box=\fbox]{equation}
    \mathcal I(y_a,q,q)\sim2 e^{-\fft{3\pi i}{\tau^2}\prod_{a=1}^3\left(\{\Delta_a\}_\tau-\fft{1+\eta_1}{2}\right)}\quad(M\text{-wing}).\label{I:Cardy:M:N=2}
\end{empheq}
This is consistent with (1.2) of \cite{GonzalezLezcano:2020yeb}, including the factor of 2.

\subsubsection*{The Cardy-like limit in the $W$-wing}
In the $W$-wing, we can simplify (\ref{eq:SCI:BA:N=2}) with $p=q$ as
\begin{equation}
\begin{split}
    \mathcal I(y_a,q,q)&=4\left(\mathcal I_{\{2,1,0\}}+\mathcal I_{\{1,2,1\}}\right)\left(1+\fft{\mathcal I_{\{1,2,0\}}}{\mathcal I_{\{2,1,0\}}+\mathcal I_{\{1,2,1\}}}\right)\\
    &\sim 4\left(\mathcal I_{\{2,1,0\}}+\mathcal I_{\{1,2,1\}}\right).
\end{split}\label{I:Cardy:W:N=2}
\end{equation}
Since $\mathcal I_{\{2,1,0\}}$ and $\mathcal I_{\{1,2,1\}}$ have the same exponential leading order in the Cardy-like limit, we must keep track of both contributions to evaluate the SU(2) index. We compute their sum in two different cases: $\eta_1=-\eta_2$ and $\eta_1=\eta_2$. Recall that $\eta_C\in\{\pm1\}$ from (\ref{eq:eta}) so these are the only options.

First, when $\eta_1=-\eta_2$, substituting (\ref{I:210:Cardy}) and (\ref{I:121:Cardy}) into (\ref{I:Cardy:W:N=2}) simply gives
\begin{empheq}[box=\fbox]{equation}
\begin{split}
	\mathcal I(y_a,q,q)&\sim-\fft{1}{2\sqrt2}e^{-\fft{\pi i}{2\tau^2}\prod_{a=1}^3\left(\{2\Delta_a\}_\tau-\fft{1+\eta_2}{2}\right)+\fft{\pi i}{\tau^2}\prod_{a=1}^3\left(\{\Delta_a\}_\tau-\fft{1+\eta_1}{2}\right)}\\
	&\quad~\,(W\text{-wing},~\eta_1=-\eta_2).
\end{split}\label{I:Cardy:W:N=2:1=-2}
\end{empheq}

For $\eta_1=\eta_2$, substituting (\ref{I:210:Cardy}) and (\ref{I:121:Cardy}) into (\ref{I:Cardy:W:N=2}) gives
\begin{equation}
\begin{split}
	\mathcal I(y_a,q)&\sim\fft{X^\text{SU(2)}}{4}e^{-\fft{\pi i}{2\tau^2}\prod_{a=1}^3\left(\{2\Delta_a\}_\tau-\fft{1+\eta_2}{2}\right)+\fft{\pi i}{\tau^2}\prod_{a=1}^3\left(\{\Delta_a\}_\tau-\fft{1+\eta_1}{2}\right)+\fft{(6-5\eta_1)\pi i}{12}}\\
	&\quad\times\prod_{a=1}^3\fft{\psi(\fft{\{\Delta_a\}_\tau}{\tau}-1)}{\psi(\fft{1-\{\Delta_a\}_\tau}{\tau}+1)}\qquad(W\text{-wing},~\eta_1=\eta_2),
\end{split}\label{I:Cardy:W:N=2:X}
\end{equation}
where $X^\text{SU(2)}$ is a complicated function of chemical potentials defined in (\ref{eq:X:SU(2)}). Following Appendix \ref{App:Cardy:N=2}, we can approximate $X^\text{SU(2)}$ as
\begin{equation}
    X^\text{SU(2)}\sim\fft{4\Delta^\text{SU(2)}}{\tau}+2\eta_1-\fft{2i}{\pi},\label{eq:X:SU(2):approx}
\end{equation}
where we have introduced $\Delta^\text{SU(2)}$ as
\begin{equation}
	\Delta^\text{SU(2)}=\begin{cases}
	\{1/2+\Delta_3\}_\tau & (\eta_1=\eta_2=-1)\\
	1-\{1/2+\Delta_1\}_\tau & (\eta_1=\eta_2=1)
	\end{cases},\label{Delta:SU(2)}
\end{equation}
under the ordering (without loss of generality)
\begin{equation}
	0<\{\tilde\Delta_1\}<\{\tilde\Delta_2\}<\{\tilde\Delta_3\}<1.
\end{equation}
Refer to (\ref{u:component}) for the definition of $\tilde\Delta_a$. Substituting (\ref{eq:X:SU(2):approx}) back into (\ref{I:Cardy:W:N=2:X}) then gives
\begin{empheq}[box=\fbox]{equation}
\begin{split}
	&\mathcal I(y_a,q,q)\\
	&\sim\bigg(\fft{\Delta^\text{SU(2)}}{\tau}+\fft{\eta_1}{2}-\fft{i}{2\pi}\bigg)e^{-\fft{\pi i}{2\tau^2}\prod_{a=1}^3\left(\{2\Delta_a\}_\tau-\fft{1+\eta_2}{2}\right)+\fft{\pi i}{\tau^2}\prod_{a=1}^3\left(\{\Delta_a\}_\tau-\fft{1+\eta_1}{2}\right)+\fft{(6-5\eta_1)\pi i}{12}}\\
	&\quad~\,(W\text{-wing},~\eta_1=\eta_2).
\end{split}\label{I:Cardy:W:N=2:1=2}
\end{empheq}
As we have explained, for the configuration of chemical potentials satisfying $\eta_1=\eta_2$ within the W-wing, the Jacobian of the 2-center BAE solutions vanishes in the leading Cardy-like limit. This situation forces us to keep track of  the first exponentially suppressed term in the Cardy-like expansion.  For the configuration of chemical potentials satisfying $\eta_1=-\eta_2$ within the W-wing, however, the Jacobian of the 2-center BAE solutions does not vanish in the leading Cardy-like limit and affords us the possibility of neglecting the first exponentially suppressed terms.

\subsection{Numerical investigation}\label{sec:SU(2):num}
Thus far we have discussed the treatment of the SU(2) SCI using the BA approach. In this subsection we want to confront our analysis with the  full index which can be obtained by direct integration in the case of small rank $N$. The integral expression of the $\mathcal N=4$ SU($N$) SCI (\ref{SCI:N=4:integral}) reduces to a one-dimensional integral for the $N=2$ case as 
\begin{equation}
    \mathcal I(y_a,p,q)=\fft{(p;p)_\infty(q;q)_\infty}{2}\prod_{a=1}^3\Gamma(y_a;p,q)\oint\fft{dz_1}{2\pi iz_1}\fft{\prod_{a=1}^3\Gamma(z_1^2y_a;p,q)\Gamma(z_1^{-2}y_a;p,q)}{\Gamma(z_1^2;p,q)\Gamma(z_1^{-2};p,q)}.\label{I:SU(2):num}
\end{equation}
We can obtain the SU(2) index directly by evaluating the integral (\ref{I:SU(2):num}) numerically. Taking the numerical answer as reference, we will confirm in several examples that the numerical integral matches the analytic result for $\mathcal I(y_a,p,q)$ from the BA formula (\ref{eq:SCI:BA:N=2}) in both asymptotic regions discussed in the two previous subsections. The comparison with the numerical evaluation of the index has the added bonus of showing us where each approximation breaks down. 

\begin{figure}[t]
	\centering
	\includegraphics[scale=.39]{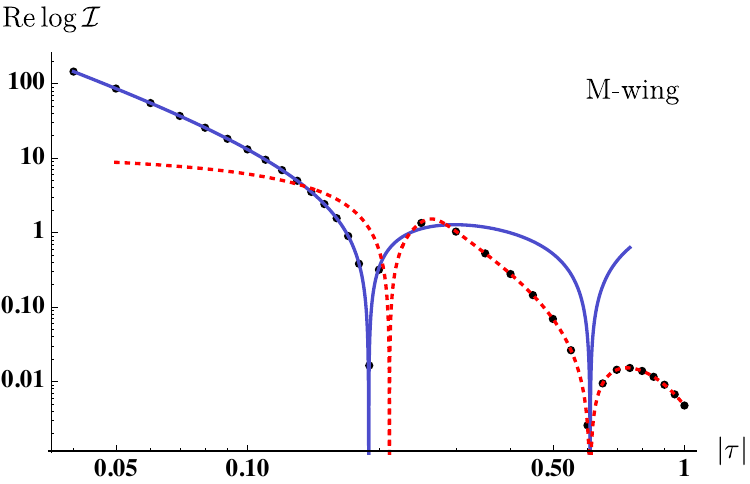}
	\includegraphics[scale=.39]{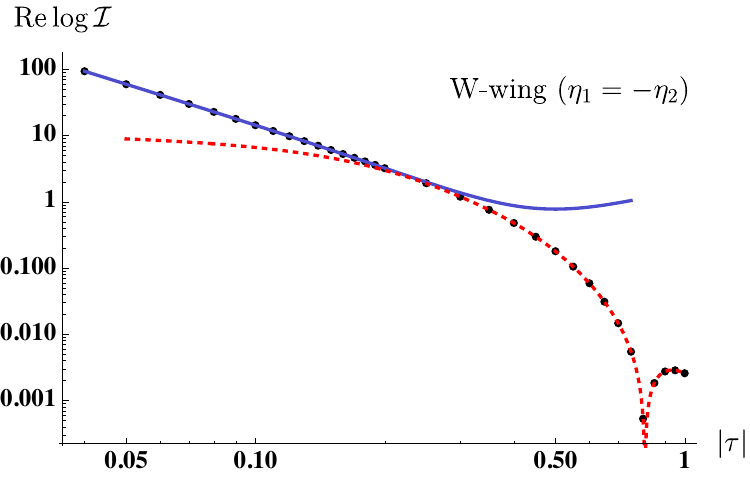}
	\includegraphics[scale=.39]{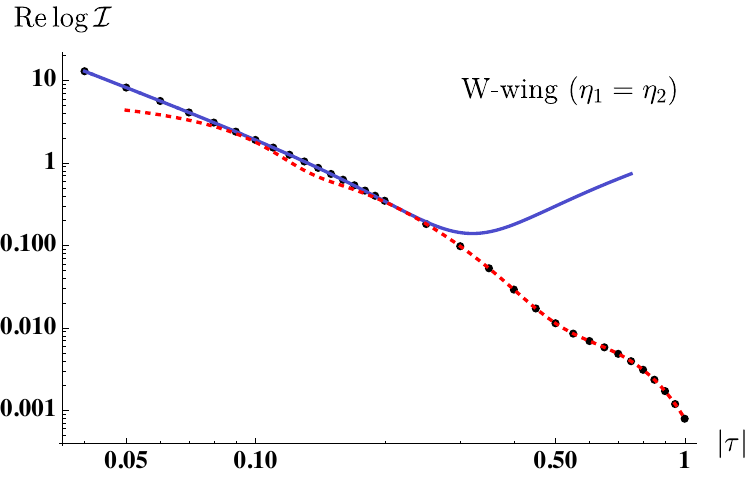}
	\caption{Plots of $\Re\log\mathcal I(y_a,q,q)$ versus $|\tau|$ for $q=e^{2\pi i\tau}$ where $\tau$ has the phase $\tau=|\tau|e^{\fft{2\pi i}{3}}$. From the left to the right, we have chosen $\Delta_a=\fft13+\fft{2\tau}{3}$, $\Delta_a=\fft23+\fft{2\tau}{3}$, and $\Delta_a=\{\fft45+\fft{2\tau}{3},\fft45+\fft{2\tau}{3},\fft25+\fft{2\tau}{3}\}$ respectively. The black dots are from numerical evaluation of the integral (\ref{I:SU(2):num}) and the dotted red lines are from the low-temperature expansion (\ref{I:low:generic}) up to order $x^{30}$. The blue lines corresponds to the Cardy-like expansions (\ref{I:Cardy:M:N=2}), (\ref{I:Cardy:W:N=2:1=-2}), and (\ref{I:Cardy:W:N=2:1=2}) respectively. The above plots show that the low-temperature\,(Cardy-like) expansions are consistent with the numerical integral where $|\tau|$ is large\,(small). \label{SU(2):num:BA}}
\end{figure}

Our results are illustrated in Fig.~\ref{SU(2):num:BA}, where the black dots represent the direct numerical evaluation of (\ref{I:SU(2):num}). We compare the numerical results with two asymptotic expansions: (i) The low temperature (i.e. large $|\tau|$) expansion represented by a red dotted line and discussed in subsection \ref{LowTSU2};  (ii) The Cardy-like (i.e. small $|\tau|$) expansion represented by an solid blue line and discussed in subsection \ref{Cardy-SU2}. Here we identify $p=q\,(=h)$ with $a=b=1$ and thereby $\sigma=\tau\,(=\omega)$. Recall that the SCI is defined for $|p|,|q|<1$ so $0<\arg\tau<\pi$.
Let us summarize our findings in three main points:

\begin{itemize}
    \item The results displayed in Fig.~\ref{SU(2):num:BA} support the efficacy of the BA approach and explicitly validate that the BA formula (\ref{eq:SCI:BA}) gives the exact SCI for the SU(2) case. In particular, \underline{only the standard} BAE solutions (\ref{sol:BAE:N=2:yes}) contribute to the SU(2) index. Non-standard BAE solutions (\ref{sol:BAE:N=2:no}) do not contribute to the SU(2) index. 
    
    \item The low temperature expansion is expected to have the radius of convergence $|q|=1$. Hence (\ref{I:low:generic}) is supposed to match the numerical results for any $\tau$ with $\Im\tau>0$, provided one keeps track of as many terms as necessary. Fig. \ref{SU(2):num:BA} shows that the expansion up to order $x^{30}$ is only valid for $|\tau|\gtrsim 0.2$. This result can be improved if one adds more terms in the series expansion. See Appendix \ref{App:lowT} for some examples.
    
    \item The Cardy-like expansions (\ref{I:Cardy:M:N=2}), (\ref{I:Cardy:W:N=2:1=-2}), and (\ref{I:Cardy:W:N=2:1=2}) are valid up to exponentially suppressed terms of the form $\mathcal O(e^{-1/|\tau|})$. Hence they match the numerical results in the small $|\tau|$ region only.

\end{itemize}




\section{The SU(3) index}\label{sec:SU(3)}
In this section, we investigate the extent to which the BA formula (\ref{eq:SCI:BA}) yields the full SU(3) index following a path parallel to that followed  in the SU(2) case. For $N=3$, the BAE (\ref{eq:BAE}) reduces to  two  transcendental equations as
\begin{equation}
\begin{split}
    e^{-2\pi i\lambda}&=\prod_\Delta\fft{\theta_1(\Delta+u_{21};\omega)}{\theta_1(\Delta-u_{21};\omega)}\fft{\theta_1(\Delta+u_{31};\omega)}{\theta_1(\Delta-u_{31};\omega)}=\prod_\Delta\fft{\theta_1(\Delta-u_{21};\omega)}{\theta_1(\Delta+u_{21};\omega)}\fft{\theta_1(\Delta+u_{32};\omega)}{\theta_1(\Delta-u_{32};\omega)}\\
    &\in\{1,w,w^2\},\label{eq:BAE:N=3}
\end{split}
\end{equation}
where $w=e^{\fft{2\pi i}{3}}$ is a primitive cube root of unity and $\Delta$ take values in $\Delta\in\{\Delta_1,\Delta_2,-\Delta_1-\Delta_2\}$. Since each transcendental equation is a multi-variable function of $u_{21}$ and $u_{31}$ and the two transcendental equations are coupled, it is difficult to classify all possible BAE solutions under the identification $(u_{21},u_{31})\sim(u_{21}+\mathbb Z+\mathbb Z\omega,u_{31}+\mathbb Z+\mathbb Z\omega)$ as in the SU(2) case. The known standard BAE solutions are given as
\begin{equation}
\begin{split}
    (u_{21},u_{31})\in&\left\{(\fft13,\fft23),(\fft\omega3,\fft{2\omega}{3}),(\fft{1+\omega}{3},\fft{2(1+\omega)}{3}),(\fft{2+\omega}{3},\fft{2(2+\omega)}{3})\right\}\\
    &\cup\left\{(\fft23,\fft13),(\fft{2\omega}{3},\fft\omega3),(\fft{2(1+\omega)}{3},\fft{1+\omega}{3}),(\fft{2(2+\omega)}{3},\fft{2+\omega}{3})\right\},\label{sol:BAE:N=3}
\end{split}
\end{equation}
where they are denoted by triples of integers $\{3,1,0\}$, $\{1,3,0\}$, $\{1,3,1\}$, and $\{1,3,2\}$ respectively in the conventions of \cite{Hong:2018viz}\footnote{The solutions in the second line of (\ref{sol:BAE:N=3}) have the same three-integer notation as the ones in the first line: switching $u_{21}\leftrightarrow u_{31}$ does not change the three-integer notation.}. Note that above, the second line is a permutation of the first one. A complex 1-dimensional continuous family of non-standard BAE solutions was also found in \cite{ArabiArdehali:2019orz}. Even though its full analytic expression is  not yet  known, a special point within the family of solution is known explicitly as
\begin{equation}
    (u_{21},u_{31})\in\left\{(\fft12,\fft\omega2),(\fft\omega2,\fft12)\right\}.\label{sol:BAE:N=3:NS}
\end{equation}
For all the known $N=3$ BAE solutions (\ref{sol:BAE:N=3}) and (\ref{sol:BAE:N=3:NS}), the value of $\lambda$ is given as $e^{-2\pi i\lambda}=1$. Hence, in the SU(3) case, the value of $e^{-2\pi i\lambda}$ is not a good criteria to distinguish standard solutions and non-standard ones.

We now consider the contribution from a standard BAE solution $(u_{21},u_{31})=(u^\star,v^\star)$, representing an arbitrary element of the 8 solutions listed in (\ref{sol:BAE:N=3}), to the SCI through the BA formula (\ref{eq:SCI:BA}). Using the double-periodicity of the BA operator (\ref{Q:double:period}), we set
\begin{equation}
    (u^\star,v^\star)=(x^\star+y^\star\omega,w^\star+z^\star\omega)\qquad\text{with}\quad-1<y^\star+z^\star\leq0,~0\leq 2y^\star-z^\star<3,
\end{equation}
without loss of generality. From this BAE solution $(u_{21},u_{31})=(u^\star,v^\star)$, we can generate total 9 inequivalent elements $\{u_1,u_2,u_3\}$ within $\mathcal M_\text{BAE}$ (\ref{M:BAE}) using the properties of the BA operator (\ref{Q:properties}) as
\begin{equation}
    \mathcal M_\text{BAE}\ni\{-\fft{u^\star+v^\star}{3}+\fft{r+s_1\omega}{3},\fft{2u^\star-v^\star}{3}+\fft{r+s_2\omega}{3},\fft{-u^\star+2v^\star}{3}+\fft{r+s_3\omega}{3}\},
\end{equation}
where
\begin{equation}
\begin{split}
    r&\in\{0,1,2\},\\
    \{s_1,s_2,s_3\}&\in\begin{cases}
    \{\{0,0,0\},\{1,1,-2\},\{2,2,-4\}\} & (0\leq 2y^\star-z^\star<1)\\
    \{\{0,0,0\},\{1,1,-2\},\{2,-1,-1\}\} & (1\leq 2y^\star-z^\star<2)\\
    \{\{0,0,0\},\{1,-2,1\},\{2,-1,-1\}\} & (2\leq 2y^\star-z^\star<3)
    \end{cases}.
\end{split}
\end{equation}
Substituting these 9 elements into the BA formula (\ref{eq:SCI:BA}), we obtain the contribution from a standard BAE solution $(u_{21},u_{31})=(u^\star,v^\star)$ to the SCI. The resulting expression can be simplified further by using the properties of the building blocks (\ref{Z:properties}) and (\ref{H:properties}) as 
\begin{equation}
\begin{split}
    &\mathcal I_{\{(u_{21},u_{31})=(u^\star,v^\star)\}}(y_q,p,q)\\
    &= 9\kappa(y_a,p,q) \fft{\sum_{m_1=1}^{ab}\sum_{m_2=1}^{ab}\mathcal Z(\{-\fft{u^\star+v^\star+(m_1+m_2)\omega}{3},\fft{2u^\star-v^\star+(2m_1-m_2)\omega}{3},\fft{-u^\star+2v^\star+(-m_1+2m_2)\omega}{3}\};\Delta,a\omega,b\omega)}{H(\{-\fft{u^\star+v^\star}{3},\fft{2u^\star-v^\star}{3},\fft{-u^\star+2v^\star}{3}\};\Delta,\omega)}.\label{eq:SCI:u^star:N=3:simple}
\end{split}
\end{equation}

Finally, the standard contribution to the SU(3) index $\mathcal I_\text{standard}(y_a,p,q)$ is given as the sum of (\ref{eq:SCI:u^star:N=3:simple}) over all standard BAE solutions $(u_{21},u_{31})=(u^\star,v^\star)$ listed in (\ref{sol:BAE:N=3}). The result can be written as
\begin{empheq}[box=\fbox]{equation}
    \mathcal I(y_a,p,q)=\underbrace{18\left(\mathcal I_{\{3,1,0\}}+\mathcal I_{\{1,3,0\}}+\mathcal I_{\{1,3,1\}}+\mathcal I_{\{1,3,2\}}\right)}_{=\,\mathcal I_\text{standard}(y_a,p,q)}+\mathcal I_\text{non-standard}(y_a,p,q),\label{eq:SCI:BA:N=3}
\end{empheq}
where we have defined
\begin{subequations}
\begin{align}
    \kappa^{-1}\mathcal I_{\{3,1,0\}}&=\fft{\sum_{m_1=1}^{ab}\sum_{m_2=1}^{ab}\mathcal Z(\{-\fft{1+(m_1+m_2)\omega}{3},\fft{(2m_1-m_2)\omega}{3},\fft{1+(-m_1+2m_2)\omega}{3}\};\Delta,a\omega,b\omega)}{H(\{-\fft{u^\star+v^\star}{3},\fft{2u^\star-v^\star}{3},\fft{-u^\star+2v^\star}{3}\};\Delta,\omega)},\label{eq:SCI:BA:N=3:310}\\
    \kappa^{-1}\mathcal I_{\{1,3,0\}}&=\fft{\sum_{m_1=1}^{ab}\sum_{m_2=1}^{ab}\mathcal Z(\{-\fft{(m_1+m_2+1)\omega}{3},\fft{(2m_1-m_2)\omega}{3},\fft{(-m_1+2m_2+1)\omega}{3}\};\Delta,a\omega,b\omega)}{H(\{-\fft{u^\star+v^\star}{3},\fft{2u^\star-v^\star}{3},\fft{-u^\star+2v^\star}{3}\};\Delta,\omega)},\label{eq:SCI:BA:N=3:130}\\
    \kappa^{-1}\mathcal I_{\{1,3,1\}}&=\fft{\sum_{m_1=1}^{ab}\sum_{m_2=1}^{ab}\mathcal Z(\{-\fft{1+(m_1+m_2+1)\omega}{3},\fft{(2m_1-m_2)\omega}{3},\fft{1+(-m_1+2m_2+1)\omega}{3}\};\Delta,a\omega,b\omega)}{H(\{-\fft{u^\star+v^\star}{3},\fft{2u^\star-v^\star}{3},\fft{-u^\star+2v^\star}{3}\};\Delta,\omega)},\label{eq:SCI:BA:N=3:131}\\
    \kappa^{-1}\mathcal I_{\{1,3,2\}}&=\fft{\sum_{m_1=1}^{ab}\sum_{m_2=1}^{ab}\mathcal Z(\{-\fft{2+(m_1+m_2+1)\omega}{3},\fft{(2m_1-m_2)\omega}{3},\fft{2+(-m_1+2m_2+1)\omega}{3}\};\Delta,a\omega,b\omega)}{H(\{-\fft{u^\star+v^\star}{3},\fft{2u^\star-v^\star}{3},\fft{-u^\star+2v^\star}{3}\};\Delta,\omega)}.\label{eq:SCI:BA:N=3:132}
\end{align}\label{eq:SCI:BA:N=3:standard}%
\end{subequations}
Note that we do not have an explicit expression for $\mathcal I_\text{non-standard}(y_a,p,q)$. The issues with this non-standard contribution will determine our ability to recover the full index using the BA approach.

\subsection{Asymptotic behaviors}\label{sec:SU(3):asympt}
As in the SU(2) case, the SU(3) index (\ref{eq:SCI:BA:N=3}) with standard contribution (\ref{eq:SCI:BA:N=3:standard}) is written in terms of elliptic functions in a complicated way. Hence, in this subsection, we investigate the SU(3) index (\ref{eq:SCI:BA:N=3}) in the asymptotic regions where we have more control of the expression (\ref{eq:SCI:BA:N=3}). Going to these limiting regions and comparison with direct numerical evaluation helps us identify quantitatively how close we are able to reconstruct the full index from the given BAE solutions. In this subsection, therefore, we investigate the SU(3) index (\ref{eq:SCI:BA:N=3}) in two asymptotic regions. Namely, in the low-temperature limit ($|\omega|\to\infty$ or $|h|\to0$) and in the Cardy-like limit ($|\omega|\to0$ or $|h|\to1$) with fixed $\arg\omega$. For simplicity, here we identify $p=q=h$ with $(a,b)=(1,1)$.

\subsubsection{The low-temperature limit}\label{sec:SU(3):asympt:low}
When $|p|=|q|=|h|<1$, we can expand the SU(3) index (\ref{eq:SCI:BA:N=3}) as a series in $h$. Specializing to the case $p=q=x^3$ with $y_a=Y_ax^2$ following the convention of \cite{Murthy:2020rbd,Agarwal:2020zwm}, we find that (\ref{eq:SCI:BA:N=3:standard}) are expanded as series in $x$ as
\begin{subequations}
\begin{align}
    \mathcal I_{\{3,1,0\}}&=\fft16+\fft13\left(\fft{1}{Y_1}+\fft{1}{Y_2}+\fft{1}{Y_3}\right)x+\mathcal O(x^2),\\
    \mathcal I_{\{1,3,0\}}&=-\fft{(1-Y_1)(1-Y_2)(1-Y_3)}{162(3-Y_1-Y_2-Y_3)^2x^4}+\mathcal O(x^{-3}),\\
    \mathcal I_{\{1,3,1\}}&=-\fft{(1-wY_1)(1-wY_2)(1-wY_3)}{162w(3-wY_1-wY_2-wY_3)^2x^4}+\mathcal O(x^{-3}),\\
    \mathcal I_{\{1,3,2\}}&=-\fft{(1-w^2Y_1)(1-w^2Y_2)(1-w^2Y_3)}{162w^2(3-w^2Y_1-w^2Y_2-w^2Y_3)^2x^4}+\mathcal O(x^{-3}),
\end{align}\label{eq:SCI:BA:N=3:standard:lowT}%
\end{subequations}
where $w=e^{\fft{2\pi i}{3}}$ is a primitive cube root of unity. The above expressions might be quite involved but it is easy to note that the sum of all  standard contributions (\ref{eq:SCI:BA:N=3:standard:lowT}) still has a non-vanishing $x^{-4}$ order.  We can compare this result with the series expansion of the SU(3) index obtained from explicitly performing the holonomy integrals  in the  representation (\ref{SCI:N=4:integral}) whose result is 
\begin{equation}
\begin{split}
    \mathcal I(y_a=Y_ax^2,p=x^3,q=x^3)&=1+(Y_1^2+Y_2^2+Y_3^2+Y_1Y_2+Y_2Y_3+Y_3Y_1)x^4\\
    &\quad-2(Y_1+Y_2+Y_3)x^5+\mathcal O(x^6).\label{SCI:SU(3):lowT}
\end{split}
\end{equation}
This expansion of the exact index does not have inverse powers of $x$. Substituting (\ref{eq:SCI:BA:N=3:standard:lowT}) and (\ref{SCI:SU(3):lowT}) into (\ref{eq:SCI:BA:N=3}) then implies
\begin{equation}
    \mathcal I(y_a,p,q)\neq\mathcal I_\text{standard}(y_a,p,q).\label{eq:SCI:BA:N=3:lowT}
\end{equation}
We have found similar results for $N=4,5$ cases. {\it We conclude that the BA formula (\ref{eq:SCI:BA}) \underline{does not} yield the  complete SCI for $N\geq3$ if we take only the standard BAE solutions denoted by three integers $\{m,n,r\}$ into account.} 

The above finding is one of the main results of this manuscript as it highlights the crucial role of non-standard solutions. In particular, our result shows that the continuous family of BAE solution in $N\geq3$ cases found in \cite{ArabiArdehali:2019orz}, should be seriously considered in attempts of reproducing the full exact superconformal index. When $N=3$, in particular, we mentioned that there is a complex 1-dimensional continuous family of BAE solutions including a special point (\ref{sol:BAE:N=3:NS}). At the moment it is not obvious how to modify the BA formula (\ref{eq:SCI:BA}) to incorporate this continuous family of BAE solutions\footnote{We are grateful to A. Cabo-Bizet for discussions on this and related topics addressed in \cite{Cabo-Bizet:2020ewf} and, in particular, for his suggestion of considering equivariant integration \`a la Atiyah-Bott-Berline-Vergne  as a guiding principle.}. The main obstruction to applying the BA formula is a zero mode of the determinant $H(\{u_i\};\Delta,\omega)$. The above analysis demonstrates the importance of this issue when computing the SCI through the BA formula (\ref{eq:SCI:BA}) in the regime where the SCI allows for a series expansion with respect to fugacities.

\subsubsection{The Cardy-like limit}\label{sec:SU(3):asympt:Cardy}
Next we investigate the Cardy-like limit ($|\omega|\to0$ or $|h|\to1$) of the SU(3) index through (\ref{eq:SCI:BA:N=3}). From here on, we use $q$ and $\tau$ instead of $h$ and $\omega$ since they are the same under the identification $p=q=h$ with $(a,b)=(1,1)$. We will also use the ``$\sim$'' symbol for equations valid up to exponentially suppressed terms of the form $\mathcal O(e^{-1/|\tau|})$.

Let us start by discussing the basic solution which has  proven to be central in the Cardy-like limit. Substituting the asymptotic behavior of $\theta_1(u;\tau)$ (\ref{elliptic:theta:1:asymp}) and $\widetilde\Gamma(u;\tau)$ (\ref{elliptic:Gamma:asymp}) into (\ref{eq:SCI:BA:N=3:130}) gives the Cardy-like limit of the contribution from the basic $\{1,3,0\}$ BAE solution as
\begin{equation}
    \log\mathcal I_{\{1,3,0\}}\sim-\fft{8\pi i}{\tau^2}\prod_{a=1}^3\left(\{\Delta_a\}_\tau-\fft{1+\eta_1}{2}\right)-\log 6.\label{I:130:Cardy}
\end{equation}
Refer to (\ref{tau-modded}) and (\ref{eq:eta}) for the definitions of the $\tau$-modded value $\{\cdot\}_\tau$ and $\eta_C\in\{\pm1\}$.

For the other three BA contributions (\ref{eq:SCI:BA:N=3:310}), (\ref{eq:SCI:BA:N=3:131}), and (\ref{eq:SCI:BA:N=3:132}), we keep track of the leading exponentially suppressed terms for the same reason in the SU(2) case: the determinant $H(\{u_i\};\Delta,\tau)$ diverges for the $\eta_1=\eta_3$ case without the leading exponentially suppressed terms. Substituting the asymptotic behaviors (\ref{elliptic:theta:1:asymp}) and (\ref{elliptic:Gamma:asymp}) into (\ref{eq:SCI:BA:N=3:310}), (\ref{eq:SCI:BA:N=3:131}), and (\ref{eq:SCI:BA:N=3:132}) then gives
\begin{equation}
\begin{split}
	&\log\mathcal I_{\{3,1,0\}}\\
	&\sim-\fft{\pi i}{3\tau^2}\prod_{a=1}^3\left(\{3\Delta_a\}_\tau-\fft{1+\eta_3}{2}\right)+\fft{\pi i}{\tau^2}\prod_{a=1}^3\left(\{\Delta_a\}_\tau-\fft{1+\eta_1}{2}\right)-\log (3^3\times 3!)\\
	&\quad+\sum_{a=1}^3\left(3\log\fft{\psi(\fft{\{1/3+\Delta_a\}_\tau}{\tau}-1)}{\psi(\fft{1-\{1/3+\Delta_a\}_\tau}{\tau}+1)}+3\log\fft{\psi(\fft{\{2/3+\Delta_a\}_\tau}{\tau}-1)}{\psi(\fft{1-\{2/3+\Delta_a\}_\tau}{\tau}+1)}+2\log\fft{\psi(\fft{\{\Delta_a\}_\tau}{\tau}-1)}{\psi(\fft{1-\{\Delta_a\}_\tau}{\tau}+1)}\right)\\
	&\quad+6\log(1-e^{-\fft{2\pi i}{3\tau}})+6\log(1-e^{-\fft{4\pi i}{3\tau}})\\
	&\quad+\begin{cases}
	\fft{2\eta_1\pi i}{3} & (\eta_1=-\eta_3)\\
	\fft{\pi i(6-5\eta_1)}{6}-2\log\sum_{J=1}^2\sum_\Delta\left(\fft{e^{-\fft{2\pi i}{\tau}(1-\{\fft{J}{3}+\Delta\}_\tau)}}{1-e^{-\fft{2\pi i}{\tau}(1-\{\fft{J}{3}+\Delta\}_\tau)}}-\fft{e^{-\fft{2\pi i}{\tau}\{\fft{J}{3}+\Delta\}_\tau}}{1-e^{-\fft{2\pi i}{\tau}\{\fft{J}{3}+\Delta\}_\tau}}\right) & (\eta_1=\eta_3)
    \end{cases},
\end{split}\label{I:310:Cardy}
\end{equation}
\begin{equation}
\begin{split}
	&\log\mathcal I_{\{1,3,1\}}\\
	&\sim-\fft{\pi i}{3\tau^2}\prod_{a=1}^3\left(\{3\Delta_a\}_\tau-\fft{1+\eta_3}{2}\right)+\fft{\pi i}{\tau^2}\prod_{a=1}^3\left(\{\Delta_a\}_\tau-\fft{1+\eta_1}{2}\right)-\log(3^3\times 3!)\\
	&\quad+\sum_{a=1}^3\left(2\log\fft{\psi(\fft{\{1/3+\Delta_a\}_\tau}{\tau}-\fft23)}{\psi(\fft{1-\{1/3+\Delta_a\}_\tau}{\tau}+\fft23)}+2\log\fft{\psi(\fft{\{2/3+\Delta_a\}_\tau}{\tau}-\fft43)}{\psi(\fft{1-\{2/3+\Delta_a\}_\tau}{\tau}+\fft43)}\right.\\
	&\kern4em~\left.+\log\fft{\psi(\fft{\{2/3+\Delta_a\}_\tau}{\tau}-\fft13)}{\psi(\fft{1-\{2/3+\Delta_a\}_\tau}{\tau}+\fft13)}+\log\fft{\psi(\fft{\{1/3+\Delta_a\}_\tau}{\tau}-\fft53)}{\psi(\fft{1-\{1/3+\Delta_a\}_\tau}{\tau}+\fft53)}+2\log\fft{\psi(\fft{\{\Delta_a\}_\tau}{\tau}-1)}{\psi(\fft{1-\{\Delta_a\}_\tau}{\tau}+1)}\right)\\
	&\quad+6\log(1-e^{-2\pi i(\fft{1}{3\tau}-\fft23)})+6\log(1-e^{-2\pi i(\fft{2}{3\tau}-\fft13)})\\
	&\quad+\begin{cases}
	\fft{4\eta_1\pi i}{3} & (\eta_1=-\eta_3)\\
	\fft{\pi i(6-5\eta_1)}{6}-2\log\sum_{J=1}^2\sum_\Delta\left(\fft{e^{-\fft{2\pi i}{\tau}(1-\{\fft{J}{3}+\Delta\}_\tau-\fft{J\tau}{3})}}{1-e^{-\fft{2\pi i}{\tau}(1-\{\fft{J}{3}+\Delta\}_\tau-\fft{J\tau}{3})}}-\fft{e^{-\fft{2\pi i}{\tau}(\{\fft{J}{3}+\Delta\}_\tau+\fft{J\tau}{3})}}{1-e^{-\fft{2\pi i}{\tau}(\{\fft{J}{3}+\Delta\}_\tau+\fft{J\tau}{3})}}\right) & (\eta_1=\eta_3)
	\end{cases},
\end{split}\label{I:131:Cardy}
\end{equation}
\begin{equation}
\begin{split}
	&\log\mathcal I_{\{1,3,2\}}\\
	&\sim-\fft{\pi i}{3\tau^2}\prod_{a=1}^3\left(\{3\Delta_a\}_\tau-\fft{1+\eta_3}{2}\right)+\fft{\pi i}{\tau^2}\prod_{a=1}^3\left(\{\Delta_a\}_\tau-\fft{1+\eta_1}{2}\right)-\log(3^3\times 3!)\\
	&\quad+\sum_{a=1}^3\left(2\log\fft{\psi(\fft{\{2/3+\Delta_a\}_\tau}{\tau}-\fft23)}{\psi(\fft{1-\{2/3+\Delta_a\}_\tau}{\tau}+\fft23)}+2\log\fft{\psi(\fft{\{1/3+\Delta_a\}_\tau}{\tau}-\fft43)}{\psi(\fft{1-\{1/3+\Delta_a\}_\tau}{\tau}+\fft43)}\right.\\
	&\kern4em~\left.+\log\fft{\psi(\fft{\{1/3+\Delta_a\}_\tau}{\tau}-\fft13)}{\psi(\fft{1-\{1/3+\Delta_a\}_\tau}{\tau}+\fft13)}+\log\fft{\psi(\fft{\{2/3+\Delta_a\}_\tau}{\tau}-\fft53)}{\psi(\fft{1-\{2/3+\Delta_a\}_\tau}{\tau}+\fft53)}+2\log\fft{\psi(\fft{\{\Delta_a\}_\tau}{\tau}-1)}{\psi(\fft{1-\{\Delta_a\}_\tau}{\tau}+1)}\right)\\
	&\quad+6\log(1-e^{-2\pi i(\fft{2}{3\tau}-\fft23)})+6\log(1-e^{-2\pi i(\fft{1}{3\tau}-\fft13)})\\
	&\quad+\begin{cases}
	\fft{4\eta_1\pi i}{3} & (\eta_1=-\eta_3)\\
	\fft{\pi i(6-5\eta_1)}{6}-2\log\sum_{J=1}^2\sum_\Delta\left(\fft{e^{-\fft{2\pi i}{\tau}(1-\{\fft{J}{3}+\Delta\}_\tau+\fft{J\tau}{3})}}{1-e^{-\fft{2\pi i}{\tau}(1-\{\fft{J}{3}+\Delta\}_\tau+\fft{J\tau}{3})}}-\fft{e^{-\fft{2\pi i}{\tau}(\{\fft{J}{3}+\Delta\}_\tau-\fft{J\tau}{3})}}{1-e^{-\fft{2\pi i}{\tau}(\{\fft{J}{3}+\Delta\}_\tau-\fft{J\tau}{3})}}\right) & (\eta_1=\eta_3)
	\end{cases}.
\end{split}\label{I:132:Cardy}
\end{equation}
Refer to Appendix \ref{App:Cardy:N=3} for details. As in the SU(2) case, the BA contributions (\ref{I:310:Cardy}), (\ref{I:131:Cardy}), and (\ref{I:132:Cardy}) have the same $\fft{1}{\tau^2}$-leading order but their sub-leading terms are different. This difference will play an important role in estimating the Cardy-like asymptotics of the SU(3) index in the $W$-wing \eqref{Wings}. 

Now, substituting (\ref{I:130:Cardy}), (\ref{I:310:Cardy}), (\ref{I:131:Cardy}), and (\ref{I:132:Cardy}) into (\ref{eq:SCI:BA:N=3}), we obtain the Cardy-like limit of the standard contribution $\mathcal I_\text{standard}(y_a,q,q)$ to the SU(3) index. Note that this contribution may not  match the Cardy-like limit of the SU(3) index because the non-standard contribution $\mathcal I_\text{non-standard}(y_q,q,q)$ may affect the result: we have already seen that this is truly the case in the low-temperature ($|\tau|\to\infty$) regime. For now, we focus on the Cardy-like limit of the standard contribution $\mathcal I_\text{standard}(y_a,q,q)$ in the $M$-wing and in the $W$-wing classified as (\ref{Wings}). 

\subsubsection*{The Cardy-like limit in the $M$-wing}
In the $M$-wing, we can simplify $\mathcal I_\text{standard}(y_a,p,q)$ in (\ref{eq:SCI:BA:N=3}) with $p=q$ as
\begin{equation}
    \mathcal I_\text{standard}(y_a,q,q)=18\mathcal I_{\{1,3,0\}}\left(1+\fft{\mathcal I_{\{3,1,0\}}+\mathcal I_{\{1,3,1\}}+\mathcal I_{\{1,3,2\}}}{\mathcal I_{\{1,3,0\}}}\right)\sim 18\mathcal I_{\{1,3,0\}}.
\end{equation}
The standard contribution is then given from (\ref{I:130:Cardy}) as
\begin{empheq}[box=\fbox]{equation}
    \mathcal I_\text{standard}(y_a,q,q)\sim3e^{-\fft{8\pi i}{\tau^2}\prod_{a=1}^3\left(\{\Delta_a\}_\tau-\fft{1+\eta_1}{2}\right)}\quad(M\text{-wing}).\label{I:Cardy:M:N=3}
\end{empheq}
This is consistent with (1.2) of \cite{GonzalezLezcano:2020yeb}, whose logarithm matches the entropy function of the  dual supersymmetric, rotating, electrically charged black hole upon the Legendre transformation with respect to chemical potentials \cite{Choi:2018hmj,Benini:2018ywd}. 

\subsubsection*{The Cardy-like limit in the $W$-wing}
In the $W$-wing, we can simplify $\mathcal I_\text{standard}(y_a,p,q)$ in (\ref{eq:SCI:BA:N=3}) with $p=q$ as
\begin{equation}
\begin{split}
    \mathcal I_\text{standard}(y_a,q,q)&=18(\mathcal I_{\{3,1,0\}}+\mathcal I_{\{1,3,1\}}+\mathcal I_{\{1,3,2\}})\left(1+\fft{\mathcal I_{\{1,3,0\}}}{\mathcal I_{\{3,1,0\}}+\mathcal I_{\{1,3,1\}}+\mathcal I_{\{1,3,2\}}}\right)\\
    &\sim 18(\mathcal I_{\{3,1,0\}}+\mathcal I_{\{1,3,1\}}+\mathcal I_{\{1,3,2\}}).
\end{split}\label{I:Cardy:W:N=3}
\end{equation}
Since $\mathcal I_{\{3,1,0\}}$, $\mathcal I_{\{1,3,1\}}$, and $\mathcal I_{\{1,3,2\}}$ have the same exponential leading order in the Cardy-like limit, we must keep track of all of them to evaluate the SU(3) index. We compute their sum in two different cases: $\eta_1=-\eta_3$ and $\eta_1=\eta_3$. Recall that $\eta_C\in\{\pm1\}$ from (\ref{eq:eta}) so these are the only options.

First, when $\eta_1=-\eta_3$, substituting (\ref{I:310:Cardy}), (\ref{I:131:Cardy}), and (\ref{I:132:Cardy}) into (\ref{I:Cardy:W:N=3}) simply gives
\begin{empheq}[box=\fbox]{equation}
\begin{split}
	\mathcal I_\text{standard}(y_a,q,q)&\sim\fft{e^{\fft{2\eta_1\pi i}{3}}+2e^{\fft{4\eta_1\pi i}{3}}}{9}e^{-\fft{\pi i}{3\tau^2}\prod_{a=1}^3\left(\{3\Delta_a\}_\tau-\fft{1+\eta_3}{2}\right)+\fft{\pi i}{\tau^2}\prod_{a=1}^3\left(\{\Delta_a\}_\tau-\fft{1+\eta_1}{2}\right)}\\
	&\quad~\,(W\text{-wing},~\eta_1=-\eta_3).
\end{split}\label{I:Cardy:W:N=3:1=-3}
\end{empheq}

For $\eta_1=\eta_3$, substituting (\ref{I:310:Cardy}), (\ref{I:131:Cardy}), and (\ref{I:132:Cardy}) into (\ref{I:Cardy:W:N=3}) gives
\begin{equation}
\begin{split}
	\mathcal I_\text{standard}(y_a,q,q)&\sim\fft{X^\text{SU(3)}}{9}e^{-\fft{\pi i}{3\tau^2}\prod_{a=1}^3\left(\{3\Delta_a\}_\tau-\fft{1+\eta_3}{2}\right)+\fft{\pi i}{\tau^2}\prod_{a=1}^3\left(\{\Delta_a\}_\tau-\fft{1+\eta_1}{2}\right)+\fft{\pi i(6-5\eta_1)}{6}}\\
	&\quad\times\left(\prod_{a=1}^3\fft{\psi(\fft{\{\Delta_a\}_\tau}{\tau}-1)}{\psi(\fft{1-\{\Delta_a\}_\tau}{\tau}+1)}\right)^2\qquad(W\text{-wing},~\eta_1=\eta_3)
\end{split}\label{I:Cardy:W:N=3:X}
\end{equation}
where $X^\text{SU(3)}$ is a complicated function of chemical potentials defined in (\ref{eq:X:SU(3)}). Following Appendix \ref{App:Cardy:N=3}, one can approximate $X^\text{SU(3)}$ as
\begin{equation}
	X^\text{SU(3)}\sim\fft{27(\Delta^\text{SU(3)})^2}{2\tau^2}+\fft{27\Delta^\text{SU(3)}(\eta_1\pi-i)}{2\pi\tau}+\fft{3(8\pi^2-15\eta_1\pi i-9)}{8\pi^2},\label{eq:X:SU(3):approx}%
\end{equation}
where we have introduced $\Delta^\text{SU(3)}$ as
\begin{equation}
	\Delta^\text{SU(3)}=\begin{cases}
		\{1/3+\Delta_3\}_\tau & (\eta_1=\eta_3=-1,~\{\tilde\Delta_3\}>2/3)\\
		\{2/3+\Delta_2\}_\tau & (\eta_1=\eta_3=-1,~\{\tilde\Delta_3\}<2/3)\\
		1-\{2/3+\Delta_1\}_\tau & (\eta_1=\eta_3=1,~\{\tilde\Delta_1\}<1/3)\\
		1-\{1/3+\Delta_2\}_\tau & (\eta_1=\eta_3=1,~\{\tilde\Delta_1\}>1/3)
	\end{cases},\label{Delta:SU(3)}
\end{equation}
under the ordering (without loss of generality)
\begin{equation}
	0<\{\tilde\Delta_1\}<\{\tilde\Delta_2\}<\{\tilde\Delta_3\}<1.
\end{equation}
Refer to (\ref{u:component}) for the definition of $\tilde\Delta_a$. Substituting (\ref{eq:X:SU(3):approx}) back into (\ref{I:Cardy:W:N=3:X}) then gives
\begin{empheq}[box=\fbox]{equation}
\begin{split}
	\mathcal I_\text{standard}(y_a,q,q)&\sim\left(\fft{3(\Delta^\text{SU(3)})^2}{2\tau^2}+\fft{3\Delta^\text{SU(3)}(\eta_1\pi-i)}{2\pi\tau}+\fft{8\pi^2-15\eta_1\pi i-9}{24\pi^2}\right)\\
	&\quad\times e^{-\fft{\pi i}{3\tau^2}\prod_{a=1}^3\left(\{3\Delta_a\}_\tau-\fft{1+\eta_3}{2}\right)+\fft{\pi i}{\tau^2}\prod_{a=1}^3\left(\{\Delta_a\}_\tau-\fft{1+\eta_1}{2}\right)+\fft{\pi i(6-5\eta_1)}{6}}\\
	&\quad~\,(W\text{-wing},~\eta_1=\eta_3).
\end{split}\label{I:Cardy:W:N=3:1=3}%
\end{empheq}
%

\subsection{Numerical investigation}\label{sec:SU(3):num}
\begin{figure}[t]
	\centering
	\includegraphics[scale=.48]{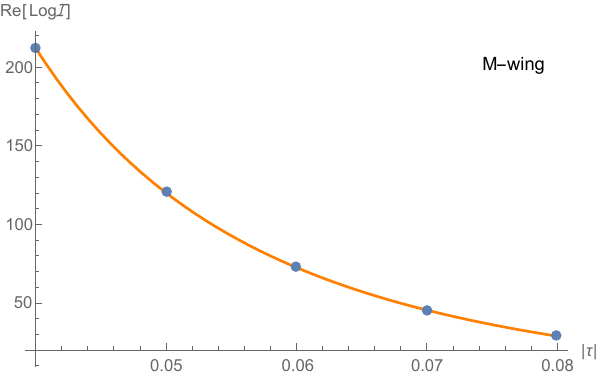}
	\includegraphics[scale=.48]{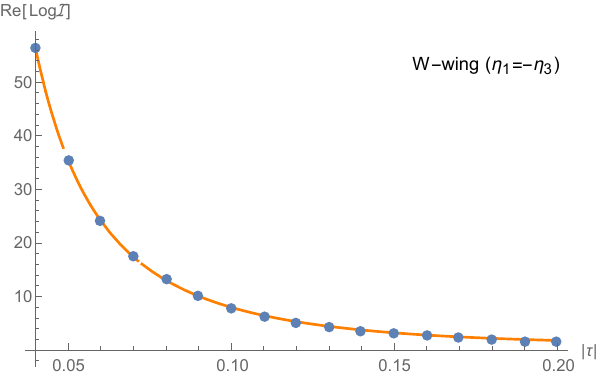}
	\includegraphics[scale=.48]{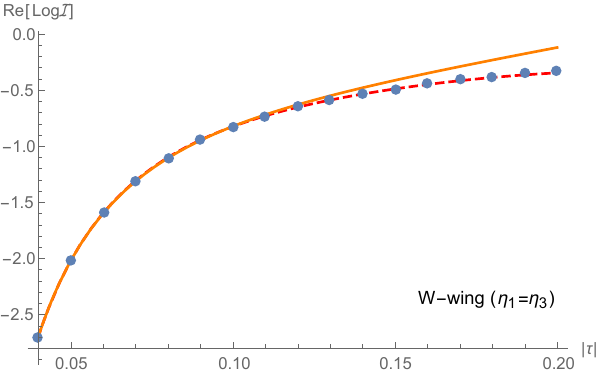}
	\caption{Plots of $\Re\log\mathcal I(y_a,q,q)$ and $\Re\log\mathcal I_\text{standard}(y_a,q,q)$ versus $|\tau|$ for $q=e^{2\pi i\tau}$. From the left to the right, we have chosen $\Delta_a=\{\fft{13}{48}+\fft{2\tau}{3},\fft{28}{48}+\fft{2\tau}{3},\fft{7}{48}+\fft{2\tau}{3}\}$, $\Delta_a=\{\fft{5}{12}+\fft{2\tau}{3},\fft{9}{12}+\fft{2\tau}{3},\fft{10}{12}+\fft{2\tau}{3}\}$, and $\Delta_a=\{\fft{3}{12}+\fft{2\tau}{3},\fft{10}{12}+\fft{2\tau}{3},\fft{11}{12}+\fft{2\tau}{3}\}$ respectively with $\tau=|\tau|e^{\fft{2\pi i}{3}}$. Blue dots are from the numerical integral (\ref{I:SU(3):num}) and orange lines are from the Cardy-like expansions (\ref{I:Cardy:M:N=3}), (\ref{I:Cardy:W:N=3:1=-3}), and (\ref{I:Cardy:W:N=3:1=3}) respectively. Red dashed line in the last plot, which matches the numerical integral better, is obtained by using $X^\text{SU(3)}$ in (\ref{eq:X:SU(3)}) without approximation. The above plots show that the Cardy-like expansion of the standard contribution $\mathcal I_\text{standard}(y_a,q,q)$ matches the numerical index $\mathcal I(y_a,q,q)$ where $|\tau|$ is small. \label{SU(3):num:BA}}
\end{figure}

The integral expression of the $\mathcal N=4$ SU($N$) SCI (\ref{SCI:N=4:integral}) reduces to a two-dimensional integral for the $N=3$ case as 
\begin{equation}
    \mathcal I(y_a,p,q)=\fft{(p;p)_\infty^2(q;q)_\infty^2}{3!}\prod_{a=1}^3\Gamma(y_a;p,q)\oint\fft{dz_1}{2\pi iz_1}\fft{dz_2}{2\pi iz_2}\prod_{i,j=1\,(i\neq  j)}^3\fft{\prod_{a=1}^3\Gamma(\fft{z_i}{z_j}y_a;p,q)}{\Gamma(\fft{z_i}{z_j};p,q)},\label{I:SU(3):num}
\end{equation}
where $z_3=1/(z_1z_2)$. We can obtain the SU(3) index directly by evaluating the integral (\ref{I:SU(3):num}) numerically. We confirmed in several examples that the numerical integral matches the analytic result for $\mathcal I_\text{standard}(y_q,p,q)$ from the BA formula (\ref{eq:SCI:BA:N=3}) in the Cardy-like limit. See Fig. \ref{SU(3):num:BA}. This means that the non-standard contribution $\mathcal I_\text{non-standard}(y_q,p,q)$ in (\ref{eq:SCI:BA:N=3}), which plays an important role in the low-temperature limit as we have observed in (\ref{eq:SCI:BA:N=3:lowT}), is suppressed in the Cardy-like limit compared to the standard one $\mathcal I_\text{standard}(y_q,p,q)$. In short, we have
\begin{equation}
    \mathcal I(y_a,q,q)\sim\mathcal I_\text{standard}(y_a,q,q)=\sum_{n|N}\sum_{r=0}^{n-1}\mathcal I_{\{N/n,n,r\}}\label{I:I-st}
\end{equation}
for the $N=3$ case. This Cardy-like asymptotics (\ref{I:I-st}) has already been anticipated for $N\leq4$ in \cite{ArabiArdehali:2019orz} by investigating a dominant holonomy configuration among various $C$-center saddles and comparing its contribution to the Cardy-like asymptotics of the SCI with numerical results: in the context of BA approach, $C$-center saddles correspond to standard BAE solutions. However, it has been shown that the Cardy-like asymptotics (\ref{I:I-st}) is \underline{not} valid for $N=5,6$ in \cite{ArabiArdehali:2019orz}. Hence for a general $N$, we expect non-standard BAE solutions affect the Cardy-like asymptotics of the SCI beyond the exponentially suppressed level. Refer to section 4.2.2 of \cite{ArabiArdehali:2019orz} for some examples of such non-standard BAE solutions.

\begin{figure}[t]
	\centering
	\includegraphics[scale=.50]{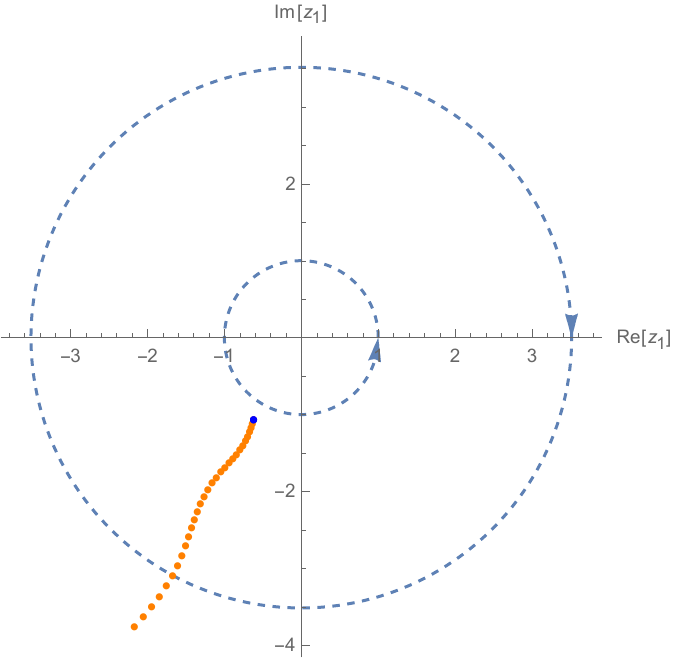}\qquad
	\includegraphics[scale=.50]{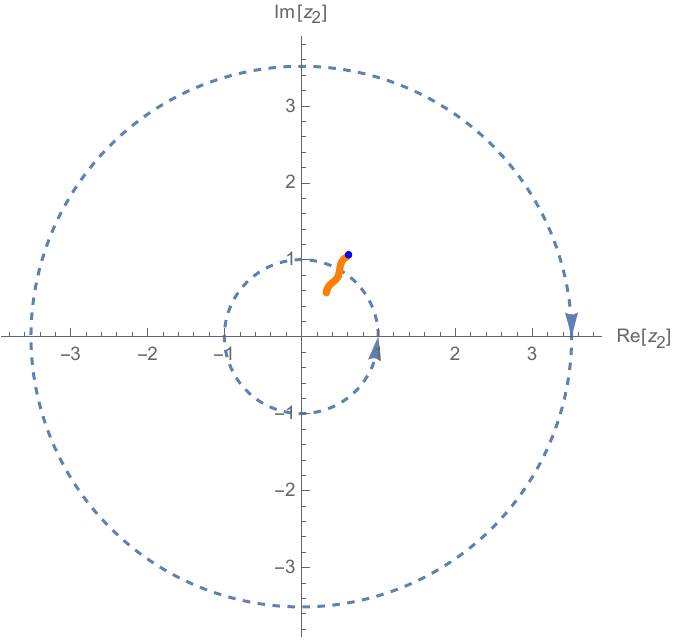}
	\caption{Numerical BAE solutions $\{z_1,z_2,z_3=1/z_1z_2\}$ of the BAE (\ref{eq:BAE:N=3}) with $\Delta_a=\{\fft15+\fft\tau4,\fft13+\fft\tau2,-\fft{8}{15}+\fft{5\tau}{4}\}$ and $\tau=1+\fft{i}{5}$. Recall $z_i=e^{2\pi iu_i}$. Blue dots denote the exact BAE solution $(u_{21},u_{32})=(\fft12,\fft\tau2)$ given in (\ref{sol:BAE:N=3:NS}) under the SU($N$) constraint $\sum_{i=1}^3u_i\in\mathbb Z$ and orange dots are numerical BAE solutions with $u_{21}=\fft12+i\fft{k}{100}~(k=1,2,\cdots,30)$. Dashed lines represent the integration contour. You may obtain different flat directions by choosing different values of $u_{21}$ and solve (\ref{eq:BAE:N=3}) numerically for $u_{31}$. \label{SU(3):num:flat}}
\end{figure}

Let us conclude this section with a numerical exploration of one non-standard solution to highlight some of its puzzling properties. We have already observed in  (\ref{eq:SCI:BA:N=3:lowT}) that  the standard contribution does not give the complete SU(3) index beyond the Cardy-like limit. We also pointed to the culprit --  a complex 1-dimensional continuous family of BAE solutions including a special point (\ref{sol:BAE:N=3:NS}). Such solution cannot be taken into account in the conventional BA formula (\ref{eq:SCI:BA}) because the formula implicitly assumes that all the BAE solutions are isolated. We explore this issue with a numerical example more explicitly. Recall that the BA formula of the index (\ref{eq:SCI:BA}) is derived from integration over the first $N-1$ holonomies along the annulus \cite{Benini:2018mlo}
\begin{equation}
    u_i:~\{0\to1\}\cup\{-\tau+1\to-\tau\}
\end{equation}
for $i=1,\cdots,N-1$. In Fig. \ref{SU(3):num:flat}, we plot the exponentiated numerical values of the first $N-1$ holonomies $z_i=e^{2\pi iu_i}~(i=1,\cdots,N-1)$ corresponding to the continuous family of BAE solutions in the $N=3$ case. It is evident that the solution is not isolated and, moreover, intersects the integration contour. These two properties clearly invalidate the conventional BA formula (\ref{eq:SCI:BA}) derived from the contour integral over the first $N-1$ holonomies along the annulus, and require a  modification that incorporates the effect of such continuous family of BAE solutions. 

\section{Discussion}\label{sec:discussion}
The BA approach to the SCI has the technical advantage of providing an exact answer for the index expressed as a sum over the solutions of the corresponding BAE. In this manuscript we have studied the details of such construction with the ultimate goal of understanding the extent to which the full index can be reconstructed from different classes of BAE solutions. Our first step was, naturally,  to group the solutions following a particular classification into standard (corresponding to a freely acting orbifold $T^2/\mathbb{Z}_m \times \mathbb{Z}_n$) and non-standard.  Since our goal is  on finite $N$ aspects we focused explicitly on $N=2$ in section  \ref{sec:SU(2)} and $N=3$ in section \ref{sec:SU(3)}; for these and other values of $N$ one can alternatively compute the index using direct numerical integration in the expression \ref{SCI:N=4:integral}.   

In section \ref{sec:SU(2)}, for SU(2),   we showed explicitly that the standard solutions (\ref{sol:BAE:N=2:yes}) were sufficient to reproduce full SCI. The non-standard solutions, presented in (\ref{NS:low-temp:N=2}), turn out, accidentally, to not contribute to the SCI. 
    
In the SU(3) case, by going to a particular regime (low temperature regime in section \ref{sec:SU(3):asympt:low}) in the SCI, we showed that the standard solutions are not enough to reproduce the index. This is the general state of affairs for $N\ge 3$. In the Cardy-like limit, we have shown that non-standard BAE solutions including dreaded family of continuous ones contribute exponentially suppressed terms to the SU(3) index at most. This supports the previous numerical investigation of \cite{ArabiArdehali:2019orz}, which implies that for $N\leq4$ the standard solutions determine the Cardy-like asymptotics of the SCI up to exponentially suppressed terms. But for $N\geq 5$, the non-standard solutions may also contribute to the Cardy-like asymptotics of the SCI beyond the exponentially suppressed level. Furthermore, we also verified, in more details than the recent analysis of \cite{GonzalezLezcano:2020yeb}, that when restricted to the $M$-wing region of fugacities, the basic solution is sufficient to reproduce the index in the Cardy-like limit up to exponentially suppressed contributions of the form ${\cal  O}(e^{-1/|\tau|})$.

One important aspect that we leave for future investigation is how to incorporate the continuous families of solutions into the expression for the BA approach to the SCI. One natural challenge is that the BA formula, as currently formulated, assumes that the solutions to the BAE are isolated; this is clearly not the case as shown in this manuscript. Another important generalization that needs to be considered is the fact that the holonomies are not all contained within a particular annulus. Indeed, in section \ref{sec:SU(3):asympt:low} we showed explicitly that the continuous family of solutions in that case intersects the integration contour, bringing in extra difficulties. There is, nevertheless, some guidance on how to generalize the BA approach to the SCI coming from equivariant integration \`a la Atiyah-Bott-Berline-Vergne, as recently discussed in \cite{Cabo-Bizet:2020ewf}.  We hope to address this issue in the future. 

One might question the need for an exact in $N$ expression for the SCI when we have demonstrated control over the leading order and, in this very manuscript, demonstrated that the non-standard solutions are exponentially suppressed in the Cardy-like limit. Our motivation is two-fold. First, we expect that such an exact in $N$ expression will help in  understanding modular properties of the full index in more details. Second, the gravitational dual of the SCI is the exact quantum entropy of the dual black holes. Namely, the exact answer in all powers of Newton's constant which will undoubtedly teach us much about quantum gravity. There are the obvious lessons from corrections to the Bekenstein-Hawking entropy recently discussed in  \cite{Liu:2017vll,Liu:2017vbl,Gang:2019uay,Benini:2019dyp,Bobev:2020egg,PandoZayas:2020iqr} in the context of AdS$_4$ balck holes. More ambitiously, is the hope that the structure of the index might guide into elucidating aspects of the putative quantum gravity path integral. For example, it would be quite interesting if there was a one-to-one correspondence between BAE solutions in field theory and gravitational configurations contributing to the path integral.

\section*{Acknowledgments}

We gratefully acknowledge useful discussions with Arash Arabi Ardehali, Alejandro Cabo-Bizet and Abhijit Gadde. This work was supported in part by the U.S. Department of Energy under grant DE-SC0007859. JH is supported in part by a Grant for Doctoral Study from the Korea Foundation for Advanced Studies.

\appendix
\section{Elliptic functions}\label{App:functions}
Here we gather some definitions and a few useful identities for elliptic functions that are used in the main body of the paper. 
\subsection{Definitions}
The Pochhammer symbol is defined as
\begin{equation}
	(z;q)_{\infty}=\prod_{k=0}^\infty(1-zq^k).\label{def:pochhammer}
\end{equation}
The elliptic theta functions have the following product forms:
\begin{subequations}
\begin{align}
	\theta_0(u;\tau)&=\prod_{k=0}^\infty(1-e^{2\pi i(u+k\tau)})(1-e^{2\pi i(-u+(k+1)\tau)}),\label{eq:theta:0}\\
	\theta_1(u;\tau)&=-ie^{\fft{\pi i\tau}{4}}(e^{\pi iu}-e^{-\pi iu})\prod_{k=1}^\infty(1-e^{2\pi ik\tau})(1-e^{2\pi i(k\tau+u)})(1-e^{2\pi i(k\tau-u)})\nn\\
	&=ie^{\fft{\pi i\tau}{4}}e^{-\pi iu}\theta_0(u;\tau)\prod_{k=1}^\infty(1-e^{2\pi ik\tau}).\label{eq:theta:1}
\end{align}\label{eq:theta}%
\end{subequations}
The elliptic Gamma function and the `tilde' elliptic Gamma function are defined as
\begin{subequations}
\begin{align}
	\Gamma(z;p,q)&=\prod_{j,k=0}^\infty\fft{1-p^{j+1}q^{k+1}z^{-1}}{1-p^jq^kz},\label{def:gamma}\\
	\widetilde\Gamma(u;\sigma,\tau)&=\prod_{j,k=0}^\infty\fft{1-e^{2\pi i[(j+1)\sigma+(k+1)\tau-u]}}{1-e^{2\pi i[j\sigma+k\tau+u]}}.\label{def:tilde:gamma}
\end{align}
\end{subequations}
For $p=q$, we abbreviate $\Gamma(z;q,q)$ and $\widetilde\Gamma(u;\tau,\tau)$ as $\Gamma(z,q)$ and $\widetilde\Gamma(u;\tau)$ respectively. We also define a special function $\psi(u)$ as
\begin{equation}
	\psi(u)\equiv\exp[u\log(1-e^{-2\pi iu})-\fft{1}{2\pi i}\text{Li}_2(e^{-2\pi iu})].\label{def:psi}
\end{equation}
The $\psi$-function satisfies
\begin{subequations}
\begin{align}
    \log\psi(u)&=\sum_{n=1}^\infty\fft{i-2\pi nu}{2\pi n^2}e^{-2\pi inu}\quad(\Im u<0),\label{psi:asymp}\\
    \psi(u+1)&=(1-e^{-2\pi iu})\psi(u).\label{psi:period}
\end{align}
\end{subequations}
%

\subsection{Basic properties}
The elliptic theta functions have quasi-double-periodicity, namely
\begin{subequations}
\begin{align}
	\theta_0(u+m+n\tau;\tau)&=(-1)^ne^{-2\pi inu}e^{-\pi in(n-1)\tau}\theta_0(u;\tau),\label{theta0:periodic}\\
	\theta_1(u+m+n\tau;\tau)&=(-1)^{m+n}e^{-2\pi inu}e^{-\pi in^2\tau}\theta_1(u;\tau),\label{theta1:periodic}
\end{align}
\end{subequations}
for $m,n\in\mathbb Z$. The inversion formula of $\theta_0(u;\tau)$ can be written simply as
\begin{equation}
    \theta_0(-u;\tau)=-e^{-2\pi iu}\theta_0(u;\tau)\label{theta0:inversion}.
\end{equation}

The elliptic Gamma function also has quasi-double-periodicity, namely
\begin{equation}
	\widetilde\Gamma(u;\sigma,\tau)=\widetilde\Gamma(u+1;\sigma,\tau)=\theta_0(u;\tau)^{-1}\widetilde\Gamma(u+\sigma;\sigma,\tau)=\theta_0(u;\sigma)^{-1}\widetilde\Gamma(u+\tau;\sigma,\tau).\label{Gamma:periodic}
\end{equation}
It also satisfies the inversion formula
\begin{equation}
    \widetilde\Gamma(u;\sigma,\tau)=\widetilde\Gamma(\sigma+\tau-u;\sigma,\tau)^{-1}.\label{Gamma:inversion}
\end{equation}
The following identity in \cite{Benini:2018mlo} is also useful:
\begin{equation}
\begin{split}
    \widetilde\Gamma(u+mab\omega;a\omega,b\omega)&=(-e^{2\pi iu})^{-\fft{abm^2}{2}+\fft{m(a+b-1)}{2}}(e^{2\pi i\omega})^{-\fft{abm^3}{6}+\fft{ab(a+b)m^2}{4}-\fft{(a^2+b^2+3ab-1)m}{12}}\\
    &\quad\times\theta_0(u;\omega)^m\widetilde\Gamma(u;a\omega,b\omega).\label{Gamma:periodic:2}
\end{split}
\end{equation}
%

\subsection{Asymptotic behaviors}
For small $|\tau|$ with fixed $0<\arg\tau<\pi$, the Pochhammer symbol can be approximated as
\begin{equation}
	\log(q;q)_\infty=-\fft{\pi i}{12}(\tau+\fft{1}{\tau})-\fft12\log(-i\tau)+\mathcal O(e^{-\fft{2\pi\sin(\arg\tau)}{|\tau|}}).\label{pochhammer:asymp}
\end{equation}

To study asymptotic behaviors of elliptic functions, first we introduce a $\tau$-modded value of a complex number $u$, namely $\{u\}_\tau$, as
\begin{equation}
	\{u\}_\tau\equiv u-\lfloor\Re u-\cot(\arg\tau)\Im u\rfloor\quad(u\in\mathbb C).\label{tau-modded}
\end{equation}
By definition, the $\tau$-modded value satisfies 
\begin{equation}
	\{u\}_\tau=\{\tilde u\}_\tau+\check u\tau,\qquad
	\{-u\}_\tau=\begin{cases}
	1-\{u\}_\tau & (\tilde u\notin\mathbb Z)\\
	-\{u\}_\tau & (\tilde u\in\mathbb Z),
	\end{cases}
\end{equation}
where we have defined $\tilde u,\check u\in\mathbb R$ as 
\begin{equation}
	u=\tilde u+\check u\tau.\label{u:component}
\end{equation}
Note that, for a real number $x$, a $\tau$-modded value $\{x\}_\tau$ reduces to a normal modded value $\{x\}$ defined as 
\begin{equation}
	\{x\}\equiv x-\lfloor x\rfloor\quad(x\in\mathbb R).\label{modded}
\end{equation}
Bernoulli functions $B_n(\cdot)$ satisfy the following useful identity written in terms of the $\tau$-modded value: 
\begin{equation}
	\sum_{J=0}^{C-1}B_n(\{\fft{J}{C}+u\}_\tau)=\fft{1}{C^{n-1}}B_n(\{Cu\}_\tau).\label{Bernoulli:identity}
\end{equation}

Now, the asymptotic behavior of elliptic functions for a small $|\tau|$ with fixed $0<\arg\tau<\pi$ are given as follows: 
\begin{equation}
\begin{split}
	\log\theta_0(u;\tau)&=\fft{\pi i}{\tau}\{u\}_\tau(1-\{u\}_\tau)+\pi i\{u\}_\tau-\fft{\pi i}{6\tau}(1+3\tau+\tau^2)\\
	&\quad+\log(1-e^{-\fft{2\pi i}{\tau}(1-\{u\}_\tau)})\left(1-e^{-\fft{2\pi i}{\tau}\{u\}_\tau}\right)+\mathcal O(e^{-\fft{2\pi\sin(\arg\tau)}{|\tau|}}),\label{elliptic:theta:0:asymp}
\end{split}
\end{equation}
\begin{equation}
\begin{split}
	\log\theta_1(u;\tau)&=\fft{\pi i}{\tau}\{u\}_\tau(1-\{u\}_\tau)-\fft{\pi i}{4\tau}(1-\tau)+\pi i\lfloor\Re u-\cot(\arg\tau)\Im u\rfloor-\fft12\log\tau\\
	&\quad+\log(1-e^{-\fft{2\pi i}{\tau}(1-\{u\}_\tau)})\left(1-e^{-\fft{2\pi i}{\tau}\{u\}_\tau}\right)+\mathcal O(e^{-\fft{2\pi\sin(\arg\tau)}{|\tau|}}),\label{elliptic:theta:1:asymp}
\end{split}
\end{equation}
\begin{equation}
\begin{split}
	\log\widetilde\Gamma(u;\tau)&=2\pi i\,Q(\{u\}_\tau;\tau)-\log\psi(\fft{\{u\}_\tau}{\tau}-1)-\log\psi(\fft{1-\{u\}_\tau}{\tau}+1)+\mathcal O(e^{-\fft{2\pi\sin(\arg\tau)}{|\tau|}}),\\
	Q(u;\tau)&\equiv-\fft{B_3(u)}{6\tau^2}+\fft{B_2(u)}{2\tau}-\fft{5}{12}B_1(u)+\fft{\tau}{12}.\label{elliptic:Gamma:asymp}
\end{split}
\end{equation}
%

\section{Series expansion of the SU(2) index}\label{App:lowT}

For finite $N$, the elliptic hypergeometric integral representation, (\ref{SCI:N=4:integral}), leads to a direct evaluation of the index by explicit integration.  While the elliptic Gamma function is not elementary, its product representation, (\ref{def:gamma}), allows for a series expansion of the finite-$N$ index.  This was explicitly realized in \cite{Murthy:2020rbd,Agarwal:2020zwm}, where the series expansion of the $\mathcal N=4$ U($N$) index was investigated for finite $N$ with the simplest possible configuration of fugacities, namely $p=q=y_a^{3/2}$.

For even moderately large values of $N\gtrsim10$, the $(N-1)$-dimensional integral soon becomes computationally expensive.  However, the $N=2$ case can readily be pushed to fairly high order in the series expansion.  Here we have explicitly
\begin{align}
    \mathcal I^{\mathrm{SU}(2)}(y_a,p,q)&=\fft{(p;p)_\infty(q;q)_\infty\Pi\Gamma(y_a;p,q)}2\int_0^1du\,W(z;p;q)\Pi\Gamma(zy_a;p,q)\Pi\Gamma(z^{-1}y_a;p,q),\nn\\
    \mathcal I^{\mathrm{U}(2)}(y_a,p,q)&=\fft{\left((p;p)_\infty(q;q)_\infty\Pi\Gamma(y_a;p,q)\right)^2}2\int_0^1du\,W(z;p;q)\Pi\Gamma(zy_a;p,q)\Pi\Gamma(z^{-1}y_a;p,q),
\label{eq:IU2SU2}
\end{align}
where $z=e^{2\pi iu}$ and we have defined
\begin{align}
W(z;p;q)&\equiv\fft1{\Gamma(z;p,q)\Gamma(z^{-1};pq)}=(z;p)_\infty(z^{-1}p;p)_\infty(z;q)_\infty(z^{-1}q;q)_\infty,\nn\\
\Pi\Gamma(y_a;p,q)&\equiv\prod_{a=1}^3\Gamma(y_a;p,q).
\end{align}
Using the product form of the elliptic Gamma function, (\ref{def:gamma}), we can obtain the product representation
\begin{equation}
    \Pi\Gamma(\zeta y_a;p;q)=\prod_{j,k=0}^\infty\fft{1-\zeta^{-1}Wp^{j+1}q^{k+1}+\zeta^{-2}Yp^{2j+1}q^{2k+1}-\zeta^{-3}p^{3j+2}q^{3k+2}}{1-\zeta Yp^jq^k+\zeta^2Wp^{2j+1}q^{2j+1}-\zeta^3p^{3j+1}lq^{3k+1}},
\label{eq:PiGamma}
\end{equation}
where we have defined
\begin{equation}
    Y\equiv\sum_{a=1}^3y_a,\qquad W\equiv\sum_{a=1}^3y_a^{-1}.
\end{equation}
Note that we have used the constraint $y_1y_2y_3=pq$ in deriving this expression.

We now consider the evaluation of the integral for the index, (\ref{eq:IU2SU2}).  One approach is to evaluate the integral over the holonomy as a contour integral
\begin{equation}
    \int_0^1du\quad\to\quad\oint\fft{dz}{2\pi iz},
\end{equation}
where the contour is the unit circle surrounding $z=0$ taken in the conventional direction.  This picks up the residues inside the unit circle, which can be identified from the product representation of the integrand, with the poles coming from the denominator of (\ref{eq:PiGamma}) as worked out in \cite{Goldstein:2020yvj}.  Alternatively, by truncating the infinite product at some finite order, the index becomes the integral of a rational function of the form
\begin{equation}
    \mathcal I(Y,W,p,q)\sim\int_0^1du\,f(z=e^{2\pi iu}),
\end{equation}
which just picks out the zero-mode $f_0$ in the series expansion of
\begin{equation}
    f(z)=\sum_n f_nz^n
\end{equation}
(where negative powers of $z$ are allowed).

There is still some subtlety in truncating the product representation (\ref{eq:PiGamma}), and that is that the series expansion in $\zeta$ (which corresponds to either $z$ or $z^{-1}$) will still have an infinite number of contributions at each order.  To avoid this issue, we must simultaneously expand in powers of $p$ and $q$.  This can be made explicit by introducing an expansion parameter $x$ along with a particular set of scalings of the fugacities.  The most straightforward scaling is to take
\begin{equation}
    p\to px^3,\qquad q\to qx^3,\qquad y_a\to y_ax^2,
    \label{eq:xscaling}
\end{equation}
to keep track of the orders of the series expansion with respect to $x$.  Note that this scaling is consistent with the constraint $y_1y_2y_3=pq$.

\begin{figure}[t]
\centering
\includegraphics[width=12cm]{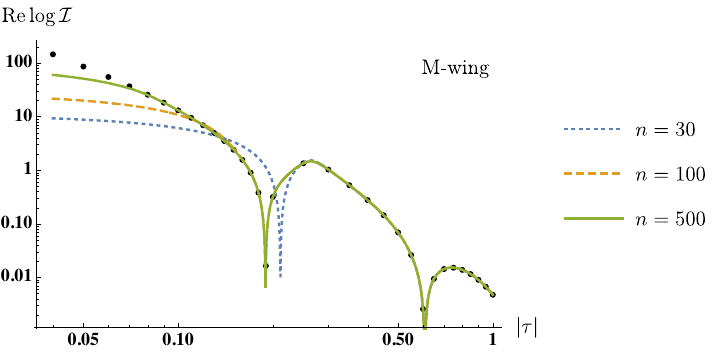}
\caption{The series evaluation of $\Re\log\mathcal I(y_a,q,q)$ compared with its numerical evaluation.  The parameters $\Delta_a=\fft13+\fft{2\tau}3$ and $\tau=|\tau|e^{\fft{2\pi i}3}$ correspond to the left panel of Fig.~\ref{SU(2):num:BA}.  The series is truncated at order $x^n$ where $q=x^3$ with $n=30$, 100 and 500 as indicated.}
\label{fig:seriestrunc}
\end{figure}

After expanding as a series in $x$, we then pick out the zero-mode of the integrand and multiply by the appropriate U(2) or SU(2) prefactor in (\ref{eq:IU2SU2}).  The result for the U(2) index is
\begin{equation}
\begin{split}
    \mathcal I^{\text{U}(2)}(Yx^2,Wx^{-2},px^3,qx^3)\kern-8em&\\
    &=1+Yx^2-(p+q)x^3+(-3pqW+2Y^2)x^4-(p+q)Yx^5\\
    &\quad+(-p^2+4pq-q^2-5pqWY+2Y^3)x^6+(p+q)(2pqW-Y^2)x^7\\
    &\quad+(5p^2q^2W^2+(-p^2+8pq-q^2)Y-11pqWY^2+3Y^4)x^8\\
    &\quad+(p+q)(-2pq+4pqWY-Y^3)x^9\\
    &\quad+((2p^3q-11p^2q^2+2pq^3)W+13p^2q^2W^2Y\\
    &\qquad~+(-2p^2+10pq-2q^2)Y^2-14pqWY^3+3Y^5)x^{10}+\mathcal O(x^{11}),\label{SCI:U(2):lowT}
\end{split}
\end{equation}
while the result for the SU(2) index is
\begin{equation}
\begin{split}
    \mathcal I^{\mathrm{SU}(2)}(Yx^2,Wx^{-2},px^3,qx^3)\kern-8em&\\
    &=1+(-pqW+Y^2)x^4-(p+q)Yx^5+pq(2-WY)x^6+(p+q)Y^2x^7\\
    &\quad+(p^2q^2W^2+(-p^2+pq-q^2)Y-3pqWY^2+Y^4)x^8\\
    &\quad+(p+q)Y(pqW-Y^2)x^9\\
    &\quad+(-p^2q^2W+2p^2q^2W^2Y+(p+q)^2Y^2-pqWY^3)x^{10}+\mathcal O(x^{11}).\label{I:low:generic}
\end{split}
\end{equation}
Here we recall the definitions
\begin{equation}
Y=y_1+y_2+y_3,\qquad W=\fft1{y_1}+\fft1{y_2}+\fft1{y^3}=\fft{y_1y_2+y_2y_3+y_3y_1}{pq}.
\end{equation}
The U(2) index reduces to the result of \cite{Murthy:2020rbd,Agarwal:2020zwm} in the equal fugacity case
\begin{equation}
    Y=3q^{2/3},\qquad W=3q^{-2/3},\qquad p=q.
    \label{eq:YWdef}
\end{equation}
(Alternatively, we can just set $Y=W=3$ and $p=q=1$ and retain $x$ as the expansion parameter used in \cite{Murthy:2020rbd,Agarwal:2020zwm}.)  The first 30 terms in the expansion of the SU(2) index are presented in Table~\ref{tbl:xcoefs}.  While the expansion gets unwieldy at high order for general $y_a$ fugacities, it simplifies considerably in special cases such as the equal fugacity case.

When $p$ and $q$ are related according to $p=h^a$ and $q=h^b$, expansion of the BA result (\ref{eq:SCI:BA:N=2}) is the most computationally efficient method for obtaining the series representation.  However, the advantage of the series expansion of the elliptic hypergeometric integral is that it applies in general even when $p$ and $q$ are unrelated.  In particular, while we assumed the scaling (\ref{eq:xscaling}), the expansion can also be rearranged as a double series in $p$ and $q$ with some corresponding scaling of the $y_a$ fugacities.

The series expansion can be viewed as an explicit realization of the Hamiltonian formulation of the index, (\ref{SCI:N=4}), with integer coefficients corresponding to the degeneracies at each order in powers of the fugacities.  As such, the series is expected to converge with fugacities inside the unit circle, namely $|p|<1$, $|q|<1$ and $|y_a|<1$.  However, this convergence can be extremely slow, so the series expansion is not particularly useful as one approaches the Cardy-like regime.  As a demonstration, we compare the numerically evaluated index with the series expansion obtained from (\ref{eq:SCI:BA:N=2}) truncated to different orders in Fig.~\ref{fig:seriestrunc}.

\afterpage{
\bgroup\small
\setlength\LTcapwidth\linewidth
\begin{longtable}[p]{l|l}
$n$&$d_2(n)$\\
\hline
\endhead
\hline
\multispan2\hbox{\strut}\\
\caption{Continued on next page}
\endfoot
\hline
\multispan2\hbox{\strut}\\
\caption{The coefficients $d_2(n)$ in the expansion of the SU(2) index $\mathcal I^{\mathrm{SU}(2)}(Yx^2,Wx^{-2},px^3,qx^3)=\sum_n d_2(n)x^n$ up to $\mathcal O(x^{30})$ where $Y$ and $W$ are defined in (\ref{eq:YWdef}). \label{tbl:xcoefs}}
\endlastfoot
0&$ 1 $\\
1&$ 0 $\\
2&$ 0 $\\
3&$ 0 $\\
4&$ Y^2-p q W $\\
5&$ Y (-p-q) $\\
6&$ 2 p q-p q W Y $\\
7&$ Y^2 (p+q) $\\
8&$ p^2 q^2 W^2+Y \left(-p^2+p q-q^2\right)-3 p q W Y^2+Y^4 $\\
9&$ p q W Y (p+q)+Y^3 (-p-q) $\\
10&$ 2 p^2 q^2 W^2 Y-p^2 q^2 W-p q W Y^3+Y^2 (p+q)^2 $\\
11&$ Y \left(-p^3-q^3\right)-p^2 q^2 W^2 (p+q)-p q W Y^2 (p+q)+Y^4 (p+q) $\\
12&$ -p^3 q^3 W^3+6 p^2 q^2 W^2 Y^2+p q W Y \left(2 p^2-7 p q+2 q^2\right)+Y^3 
\left(-2 p^2+3 p q-2 q^2\right)+p^2 q^2$\\&$\quad-5 p q W Y^4+Y^6 $\\
13&$ -p q W \left(p^3+q^3\right)+2 Y^2 \left(p^3+q^3\right)+2 p q W Y^3 
(p+q)+Y^5 (-p-q) $\\
14&$ 2 p^3 q^3 W^2+4 p^2 q^2 W^2 Y^3-2 p q W Y^2 \left(p^2+3 p q+q^2\right)+Y^4 
\left(2 p^2+3 p q+2 q^2\right)$\\&$\quad+Y \left(-p^4-3 p^3 q^3 W^3-p^3 q+2 p^2 q^2-p 
q^3-q^4\right)-p q W Y^5 $\\
15&$ 2 p q W Y \left(p^3+q^3\right)+p^2 q^2 W^2 Y^2 (p+q)+Y^3 \left(-3 p^3-p^2 
q-p q^2-3 q^3\right)-3 p q W Y^4 (p+q)$\\&$\quad+Y^6 (p+q) $\\
16&$ p^4 q^4 W^4+15 p^2 q^2 W^2 Y^4-p^2 q^2 W^2 Y \left(p^2-16 p q+q^2\right)+2 
p q W Y^3 \left(2 p^2-11 p q+2 q^2\right)$\\&$\quad+p^2 q^2 W \bigl(p^2-5 p 
q+q^2\bigr)+Y^5 \left(-2 p^2+5 p q-2 q^2\right)+Y^2 \left(2 p^4-10 p^3 q^3 
W^3+9 p^2 q^2+2 q^4\right)$\\&$\quad-7 p q W Y^6+Y^8 $\\
17&$ -3 p^2 q^2 W^2 Y^3 (p+q)-p^2 q^2 W^2 \left(p^3+2 p^2 q+2 p q^2+q^3\right)+p 
q W Y^2 \left(-2 p^3+7 p^2 q+7 p q^2-2 q^3\right)$\\&$\quad+Y^4 \left(3 p^3-p^2 q-p 
q^2+3 q^3\right)+Y \left(-p^5-3 p^3 q^2-3 p^2 q^3-q^5\right)+4 p q W Y^5 
(p+q)+Y^7 (-p-q) $\\
18&$ 14 p^3 q^3 W^2 Y^2+p^3 q^3+6 p^2 q^2 W^2 Y^5-p q W Y^4 \left(5 p^2+13 p q+5 
q^2\right)+Y^6 \left(2 p^2+3 p q+2 q^2\right)$\\&$\quad+Y \left(4 p^4 q^4 W^4+p q W 
(p+q)^2 \left(2 p^2-5 p q+2 q^2\right)\right)+p^3 q^3 W^3 \left(p^2-3 p 
q+q^2\right)$\\&$\quad+Y^3 \left(-4 p^4-10 p^3 q^3 W^3-p^3 q+2 p^2 q^2-p q^3-4 
q^4\right)-p q W Y^7 $\\
19&$ 6 p^2 q^2 W^2 Y^4 (p+q)-p q W (p+q) \left(p^2-p q+q^2\right)^2+p^2 q^2 W^2 
Y \left(-2 p^3+p^2 q+p q^2-2 q^3\right)$\\&$\quad+p q W Y^3 \left(8 p^3-5 p^2 q-5 p 
q^2+8 q^3\right)+Y^5 \left(-4 p^3+p^2 q+p q^2-4 q^3\right)$\\&$\quad+Y^2 \left(3 
p^5-p^4 q-p^3 q^3 W^3 (p+q)+5 p^3 q^2+5 p^2 q^3-p q^4+3 q^5\right)-5 p q W 
Y^6 (p+q)+Y^8 (p+q) $\\
20&$ -p^5 q^5 W^5+28 p^2 q^2 W^2 Y^6+p^2 q^2 W^2 Y^3 \left(-4 p^2+71 p q-4 
q^2\right)+p q W Y^5 \left(7 p^2-43 p q+7 q^2\right)$\\&$\quad+Y^7 \left(-2 p^2+7 p 
q-2 q^2\right)+Y^2 \left(15 p^4 q^4 W^4+p q W \left(-5 p^4+6 p^3 q-42 p^2 
q^2+6 p q^3-5 q^4\right)\right)$\\&$\quad+Y^4 \left(5 p^4-35 p^3 q^3 W^3-p^3 q+20 p^2 
q^2-p q^3+5 q^4\right)$\\&$\quad+Y \bigl(-24 p^4 q^4 W^3-(p-q)^2 \bigl(p^4+2 p^3 q+4 
p^2 q^2+2 p q^3+q^4\bigr)\bigr)$\\&$\quad+p^2 q^2 W^2 \left(p^4-p^3 q+10 p^2 q^2-p 
q^3+q^4\right)-9 p q W Y^8+Y^{10} $\\
21&$ -10 p^2 q^2 W^2 Y^5 (p+q)+p^3 q^3 W^3 \left(p^3+p^2 q+p q^2+q^3\right)+2 
p^2 q^2 W^2 Y^2 \left(p^3-7 p^2 q-7 p q^2+q^3\right)$\\&$\quad-2 p q W Y^4 \left(5 
p^3-6 p^2 q-6 p q^2+5 q^3\right)+Y^6 \left(4 p^3-2 p^2 q-2 p q^2+4 
q^3\right)+2 p^2 q^2 \left(p^3+q^3\right)$\\&$\quad+Y^3 \left(-5 p^5+2 p^4 q+4 p^3 q^3 
W^3 (p+q)-9 p^3 q^2-9 p^2 q^3+2 p q^4-5 q^5\right)$\\&$\quad+p q W Y \left(2 p^5-7 p^4 
q+6 p^3 q^2+6 p^2 q^3-7 p q^4+2 q^5\right)+6 p q W Y^7 (p+q)+Y^9 (-p-q) $\\
22&$ 5 p^5 q^5 W^4+8 p^2 q^2 W^2 Y^7+p^2 q^2 W^2 Y^4 \left(9 p^2+47 p q+9 
q^2\right)-3 p q W Y^6 \left(3 p^2+7 p q+3 q^2\right)$\\&$\quad+Y^8 \left(2 p^2+3 p 
q+2 q^2\right)+Y^3 \left(20 p^4 q^4 W^4+p q W \left(10 p^4-5 p^3 q-24 p^2 
q^2-5 p q^3+10 q^4\right)\right)$\\&$\quad+Y^5 \left(-6 p^4-21 p^3 q^3 W^3+p^3 q+5 p^2 
q^2+p q^3-6 q^4\right)$\\&$\quad+Y^2 \bigl((p+q)^2 \bigl(3 p^4-7 p^3 q+12 p^2 q^2-7 p 
q^3+3 q^4\bigr)-p^3 q^3 W^3 \left(p^2+37 p q+q^2\right)\bigr)$\\&$\quad+p^2 q^2 W 
\left(p^4+2 p^3 q-3 p^2 q^2+2 p q^3+q^4\right)$\\&$\quad+Y \left(-5 p^5 q^5 W^5-p^2 
q^2 W^2 \left(p^4-3 p^3 q-18 p^2 q^2-3 p q^3+q^4\right)\right)-p q W Y^9 $\\
23&$ -4 p^3 q^3 W^2 \left(p^3+q^3\right)+15 p^2 q^2 W^2 Y^6 (p+q)+p^2 q^2 W^2 
Y^3 \left(-10 p^3+21 p^2 q+21 p q^2-10 q^3\right)$\\*&$\quad+p q W Y^5 \left(14 p^3-17 
p^2 q-17 p q^2+14 q^3\right)+Y^7 \left(-4 p^3+3 p^2 q+3 p q^2-4 
q^3\right)$\\*&$\quad+Y^2 \left(p^4 q^4 W^4 (p+q)-p q W \left(6 p^5-15 p^4 q+14 p^3 
q^2+14 p^2 q^3-15 p q^4+6 q^5\right)\right)$\\*&$\quad+Y^4 \left(7 p^5-5 p^4 q-10 p^3 
q^3 W^3 (p+q)+11 p^3 q^2+11 p^2 q^3-5 p q^4+7 q^5\right)$\\*&$\quad+Y \bigl(-p^7-3 p^5 
q^2+2 p^4 q^3+2 p^3 q^4+2 p^3 q^3 W^3 \left(p^3+q^3\right)-3 p^2 
q^5-q^7\bigr)-7 p q W Y^8 (p+q)$\\*&$\quad+Y^{10} (p+q) $\\
24&$ p^6 q^6 W^6+45 p^2 q^2 W^2 Y^8-2 p^2 q^2 W^2 Y^5 \left(8 p^2-91 p q+8
q^2\right)+p q W Y^7 \left(11 p^2-71 p q+11 q^2\right)$\\&$\quad+Y^9 \left(-2 p^2+9 p
q-2 q^2\right)-p q \left(p^6+p^3 q^3+q^6\right)$\\&$\quad+Y^4 \bigl(70 p^4 q^4 W^4+p q
W \bigl(-12 p^4+22 p^3 q-125 p^2 q^2+22 p q^3-12 q^4\bigr)\bigr)$\\&$\quad+Y^6
\left(6 p^4-84 p^3 q^3 W^3-4 p^3 q+33 p^2 q^2-4 p q^3+6 q^4\right)+p^3 q^3
W^3 \left(p^4+2 p^3 q-14 p^2 q^2+2 p q^3+q^4\right)$\\&$\quad+Y^2 \bigl(-21 p^5 q^5
W^5-p^2 q^2 W^2 \bigl(p^4+24 p^3 q-111 p^2 q^2+24 p q^3+q^4\bigr)\bigr)$\\&$\quad+Y
\bigl(36 p^5 q^5 W^4+p q W \bigl(4 p^6-2 p^5 q+7 p^4 q^2-23 p^3 q^3+7 p^2
q^4-2 p q^5+4 q^6\bigr)\bigr)$\\&$\quad+Y^3 \bigl(-7 p^6+p^5 q-10 p^4 q^2+16 p^3
q^3-10 p^2 q^4+p^3 q^3 W^3 \left(5 p^2-164 p q+5 q^2\right)+p q^5-7
q^6\bigr)$\\&$\quad-11 p q W Y^{10}+Y^{12} $\\
25&$ -21 p^2 q^2 W^2 Y^7 (p+q)+p^2 q^2 W^2 Y^4 \left(15 p^3-44 p^2 q-44 p q^2+15
q^3\right)$\\&$\quad+p^3 q^3 W^2 Y \bigl(8 p^3-11 p^2 q-11 p q^2+8 q^3\bigr)+p q W
Y^6 \left(-17 p^3+25 p^2 q+25 p q^2-17 q^3\right)$\\&$\quad-p^4 q^4 W^4 \left(p^3+p^2
q+p q^2+q^3\right)$\\&$\quad+Y^3 \left(p q W \left(13 p^5-18 p^4 q+39 p^3 q^2+39 p^2
q^3-18 p q^4+13 q^5\right)-5 p^4 q^4 W^4 (p+q)\right)$\\&$\quad+Y^5 \left(-9 p^5+5 p^4
q+20 p^3 q^3 W^3 (p+q)-18 p^3 q^2-18 p^2 q^3+5 p q^4-9 q^5\right)$\\&$\quad+Y^2
\bigl(4 p^7+8 p^5 q^2-3 p^4 q^3-3 p^3 q^4+8 p^2 q^5+p^3 q^3 W^3 \left(p^3+18
p^2 q+18 p q^2+q^3\right)+4 q^7\bigr)$\\&$\quad-p q W \bigl(p^7+p^6 q+3 p^5 q^2-2 p^4
q^3-2 p^3 q^4+3 p^2 q^5+p q^6+q^7\bigr)+8 p q W Y^9 (p+q)+Y^{11} (-p-q)$\\&$\quad+4
Y^8 (p-q)^2 (p+q) $\\
26&$ -6 p^6 q^6 W^5+10 p^2 q^2 W^2 Y^9+p^2 q^2 W^2 Y^6 \left(25 p^2+98 p q+25
q^2\right)-p q W Y^8 \left(13 p^2+29 p q+13 q^2\right)$\\&$\quad+Y^{10} \left(2 p^2+3
p q+2 q^2\right)+Y^5 \left(56 p^4 q^4 W^4+p q W \left(23 p^4-25 p^3 q-71 p^2
q^2-25 p q^3+23 q^4\right)\right)$\\&$\quad+Y^7 \left(-7 p^4-36 p^3 q^3 W^3+5 p^3 q+11
p^2 q^2+5 p q^3-7 q^4\right)$\\&$\quad+Y^3 \bigl(p^2 q^2 W^2 \bigl(-15 p^4+25 p^3
q+111 p^2 q^2+25 p q^3-15 q^4\bigr)-35 p^5 q^5 W^5\bigr)$\\&$\quad+Y^2 \bigl(p^4 q^4
W^4 \left(p^2+71 p q+q^2\right)-p q W \bigl(9 p^6-16 p^5 q+11 p^4 q^2+43 p^3
q^3+11 p^2 q^4-16 p q^5+9 q^6\bigr)\bigr)$\\&$\quad+Y^4 \bigl(10 p^6-7 p^5 q+13 p^4
q^2+29 p^3 q^3+13 p^2 q^4-2 p^3 q^3 W^3 \left(7 p^2+69 p q+7 q^2\right)-7 p
q^5+10 q^6\bigr)$\\&$\quad+p^2 q^2 W^2 \bigl(p^6-2 p^5 q+p^4 q^2+3 p^3 q^3+p^2 q^4-2
p q^5+q^6\bigr)$\\&$\quad+Y \bigl(-p^8+6 p^6 q^6 W^6-3 p^6 q^2-4 p^5 q^3+7 p^4 q^4-4
p^3 q^5-3 p^2 q^6$\\&$\qquad+p^3 q^3 W^3 \left(2 p^4+p^3 q-36 p^2 q^2+p q^3+2
q^4\right)-q^8\bigr)-p q W Y^{11} $\\
27&$ -p^3 q^3 \left(p^3+q^3\right)+28 p^2 q^2 W^2 Y^8 (p+q)-p^4 q^4 W^3
\left(p^3+q^3\right)$\\&$\quad+p^2 q^2 W^2 Y^5 \bigl(-29 p^3+70 p^2 q+70 p q^2-29
q^3\bigr)+p q W Y^7 \left(21 p^3-34 p^2 q-34 p q^2+21 q^3\right)$\\&$\quad+Y^9
\left(-4 p^3+5 p^2 q+5 p q^2-4 q^3\right)$\\&$\quad+Y^4 \left(15 p^4 q^4 W^4 (p+q)-p q
W \left(21 p^5-41 p^4 q+45 p^3 q^2+45 p^2 q^3-41 p q^4+21
q^5\right)\right)$\\&$\quad+Y^6 \left(10 p^5-9 p^4 q-35 p^3 q^3 W^3 (p+q)+19 p^3
q^2+19 p^2 q^3-9 p q^4+10 q^5\right)$\\&$\quad+Y^2 \bigl(2 p^2 q^2 W^2 \bigl(2 p^5-17
p^4 q+8 p^3 q^2+8 p^2 q^3-17 p q^4+2 q^5\bigr)-p^5 q^5 W^5 (p+q)\bigr)$\\&$\quad+Y^3
\bigl(-8 p^7+5 p^6 q-17 p^5 q^2+2 p^4 q^3+2 p^3 q^4-17 p^2 q^5+3 p^3 q^3 W^3
\left(3 p^3-13 p^2 q-13 p q^2+3 q^3\right)$\\&$\qquad+5 p q^6-8 q^7\bigr)+p q W Y
\bigl(3 p^7-7 p^6 q+15 p^5 q^2-6 p^4 q^3-6 p^3 q^4+15 p^2 q^5-7 p q^6+3
q^7\bigr)$\\&$\quad-9 p q W Y^{10} (p+q)+Y^{12} (p+q) $\\
28&$ -p^7 q^7 W^7+19 p^6 q^6 W^4+66 p^2 q^2 W^2 Y^{10}+p^2 q^2 W^2 Y^7 \left(-36
p^2+373 p q-36 q^2\right)$\\*&$\quad+p q W Y^9 \bigl(15 p^2-107 p q+15
q^2\bigr)+Y^{11} \left(-2 p^2+11 p q-2 q^2\right)$\\*&$\quad+Y^6 \bigl(210 p^4 q^4
W^4+p q W \bigl(-28 p^4+41 p^3 q-287 p^2 q^2+41 p q^3-28
q^4\bigr)\bigr)$\\*&$\quad+Y^8 \left(7 \left(p^4-p^3 q+7 p^2 q^2-p q^3+q^4\right)-165
p^3 q^3 W^3\right)$\\*&$\quad+Y^4 \left(p^2 q^2 W^2 \left(23 p^4-67 p^3 q+506 p^2
q^2-67 p q^3+23 q^4\right)-126 p^5 q^5 W^5\right)$\\*&$\quad+W \bigl(2 p^8 q^2-2 p^6
q^4-5 p^5 q^5-2 p^4 q^6+2 p^2 q^8\bigr)$\\*&$\quad+Y^3 \bigl(p^4 q^4 W^4 \left(-6
p^2+323 p q-6 q^2\right)$\\*&$\qquad+p q W \bigl(18 p^6-29 p^5 q+44 p^4 q^2-173 p^3
q^3+44 p^2 q^4-29 p q^5+18 q^6\bigr)\bigr)$\\*&$\quad+Y^5 \bigl(-13 p^6+9 p^5 q-25
p^4 q^2+53 p^3 q^3-25 p^2 q^4+15 p^3 q^3 W^3 \left(2 p^2-37 p q+2
q^2\right)+9 p q^5-13 q^6\bigr)$\\*&$\quad+Y \bigl(-49 p^6 q^6 W^5-p^2 q^2 W^2
\bigl(p^6-12 p^5 q+10 p^4 q^2-102 p^3 q^3+10 p^2 q^4-12 p
q^5+q^6\bigr)\bigr)$\\*&$\quad+Y^2 \bigl(4 p^8-3 p^7 q+28 p^6 q^6 W^6+8 p^6 q^2+2 p^5
q^3+13 p^4 q^4+2 p^3 q^5+8 p^2 q^6$\\*&$\qquad+p^4 q^4 W^3 \left(25 p^2-272 p q+25
q^2\right)-3 p q^7+4 q^8\bigr)-13 p q W Y^{12}+Y^{14} $\\
29&$ -36 p^2 q^2 W^2 Y^9 (p+q)+2 p^2 q^2 W^2 Y^6 \left(23 p^3-55 p^2 q-55 p
q^2+23 q^3\right)$\\&$\quad-5 p q W Y^8 \bigl(5 p^3-9 p^2 q-9 p q^2+5
q^3\bigr)+Y^{10} \left(4 p^3-6 p^2 q-6 p q^2+4 q^3\right)$\\&$\quad+Y^5 \bigl(p q W
\bigl(33 p^5-51 p^4 q+95 p^3 q^2+95 p^2 q^3-51 p q^4+33 q^5\bigr)-35 p^4
q^4 W^4 (p+q)\bigr)$\\&$\quad+Y^7 \bigl(-11 p^5+11 p^4 q+56 p^3 q^3 W^3 (p+q)-25 p^3
q^2-25 p^2 q^3+11 p q^4-11 q^5\bigr)$\\&$\quad-4 p^3 q^3 W^2 \left(p^5-p^4 q+2 p^3
q^2+2 p^2 q^3-p q^4+q^5\right)$\\&$\quad+Y^3 \bigl(6 p^5 q^5 W^5 (p+q)-p^2 q^2 W^2
\left(17 p^5-50 p^4 q+97 p^3 q^2+97 p^2 q^3-50 p q^4+17
q^5\right)\bigr)$\\&$\quad+Y^2 \bigl(2 p^4 q^4 W^4 \bigl(p^3-8 p^2 q-8 p
q^2+q^3\bigr)$\\&$\qquad+2 p q W \bigl(-5 p^7+10 p^6 q-17 p^5 q^2+17 p^4 q^3+17 p^3
q^4-17 p^2 q^5+10 p q^6-5 q^7\bigr)\bigr)$\\&$\quad+Y^4 \bigl(13 p^7-10 p^6 q+27 p^5
q^2-10 p^4 q^3-10 p^3 q^4+27 p^2 q^5$\\&$\qquad+p^3 q^3 W^3 \bigl(-25 p^3+92 p^2 q+92 p
q^2-25 q^3\bigr)-10 p q^6+13 q^7\bigr)$\\&$\quad+Y \bigl(-p^9-2 p^7 q^2+p^6 q^3-2
p^5 q^4-2 p^4 q^5+p^3 q^6-2 p^2 q^7$\\&$\qquad+p^4 q^4 W^3 \left(-2 p^3+21 p^2 q+21 p
q^2-2 q^3\right)-q^9\bigr)+10 p q W Y^{11} (p+q)+Y^{13} (-p-q) $\\
30&$-p q W Y^{13}+\left(2 p^2+3 q p+2 q^2\right) Y^{12}+12 p^2 q^2 W^2 Y^{11}-p q \left(17
   p^2+37 q p+17 q^2\right) W Y^{10}$\\&$\quad+\left(-7 p^4-55 q^3 W^3 p^3+9 q p^3+17 q^2 p^2+9 q^3
   p-7 q^4\right) Y^9+p^2 q^2 \left(49 p^2+169 q p+49 q^2\right) W^2 Y^8$\\&$\quad+\left(120 p^4
   q^4 W^4+p q \left(35 p^4-57 q p^3-141 q^2 p^2-57 q^3 p+35 q^4\right) W\right)
   Y^7$\\&$\quad+\bigl(15 p^6-16 q p^5+28 q^2 p^4+51 q^3 p^3-5 q^3 \left(11 p^2+71 q p+11
   q^2\right) W^3 p^3+28 q^4 p^2-16 q^5 p+15 q^6\bigr) Y^6$\\&$\quad+\bigl(p^2 q^2 \bigl(-46
   p^4+106 q p^3+367 q^2 p^2+106 q^3 p-46 q^4\bigr) W^2-126 p^5 q^5 W^5\bigr)
   Y^5$\\&$\quad+\bigl(20 p^4 q^4 \left(p^2+17 q p+q^2\right) W^4$\\&$\qquad-p q \left(30 p^6-65 q p^5+59 q^2
   p^4+159 q^3 p^3+59 q^4 p^2-65 q^5 p+30 q^6\right) W\bigr) Y^4$\\&$\quad+\bigl(-10 p^8+8 q
   p^7+56 q^6 W^6 p^6-23 q^2 p^6-6 q^3 p^5+29 q^4 p^4-6 q^5 p^3$\\&$\qquad+q^3 \bigl(16 p^4-47 q
   p^3-325 q^2 p^2-47 q^3 p+16 q^4\bigr) W^3 p^3-23 q^6 p^2+8 q^7 p-10 q^8\bigr)
   Y^3$\\&$\quad+\bigl(2 p^2 q^2 \bigl(3 p^6-24 q p^5+11 q^2 p^4+57 q^3 p^3+11 q^4 p^2-24 q^5 p+3
   q^6\bigr) W^2$\\&$\qquad-p^5 q^5 \left(p^2+121 q p+q^2\right) W^5\bigr) Y^2$\\&$\quad+\bigl(-7 p^7 q^7
   W^7-p^4 q^4 \bigl(3 p^4+4 q p^3-74 q^2 p^2+4 q^3 p+3 q^4\bigr) W^4$\\&$\qquad+p q \bigl(4 p^8-11
   q p^7+15 q^2 p^6+4 q^3 p^5-37 q^4 p^4+4 q^5 p^3+15 q^6 p^2-11 q^7 p+4 q^8\bigr)
   W\bigr) Y$\\&$\quad+3 p^2 q^8+7 p^7 q^7 W^6+5 p^5 q^5+p^3 q^3 \bigl(p^6+4 q p^5-2 q^2 p^4-11
   q^3 p^3-2 q^4 p^2+4 q^5 p+q^6\bigr) W^3+3 p^8 q^2$
\end{longtable}
\egroup}


\section{Cardy-like expansions of the standard contributions}\label{App:Cardy}
In this appendix, we investigate the Cardy-like limit of the standard contributions to the $\mathcal N=4$ SU($N$) SCI through the BA formula (\ref{eq:SCI:BA}) for $N=2$ and $N=3$ cases respectively. In particular, since the contribution from the basic $\{1,N,0\}$ BAE solution have already been computed in the literature and given explicitly as (\ref{I:120:Cardy}) and (\ref{I:130:Cardy}) for $N=2$ and $N=3$ respectively, we focus on the other standard BAE solutions. As mentioned in the main text, we identify $p=q$ for simplicity. 

\subsection{SU(2) case}\label{App:Cardy:N=2}
For the SU(2) case, there are two remaining standard BAE solutions: $\{2,1,0\}$ and $\{1,2,1\}$. Their contributions to the SCI are given in (\ref{eq:SCI:BA:N=2:210:s=t}) and (\ref{eq:SCI:BA:N=2:121:s=t}) respectively.

First, the building blocks of (\ref{eq:SCI:BA:N=2:210:s=t}), or equivalently those of (\ref{eq:SCI:BA:N=2:210}) with $a=b=1$, can be computed using the asymptotic expansions (\ref{elliptic:theta:1:asymp}) and (\ref{elliptic:Gamma:asymp}) as
\begin{equation}
\begin{split}
    \log\kappa\mathcal Z_{\{2,1,0\}}&=-\fft{\pi i}{2\tau^2}\prod_{a=1}^3\left(\{2\Delta_a\}_\tau-\fft{1+\eta_2}{2}\right)+\fft{\pi i}{\tau^2}\prod_{a=1}^3\left(\{\Delta_a\}_\tau-\fft{1+\eta_1}{2}\right)\\
	&\quad+\fft{\pi i(6+5\eta_1-10\eta_2)}{12}-\log\tau-\log 2!\\
	&\quad+\sum_{a=1}^3\left(2\log\fft{\psi(\fft{\{1/2+\Delta_a\}_\tau}{\tau}-1)}{\psi(\fft{1-\{1/2+\Delta_a\}_\tau}{\tau}+1)}+\log\fft{\psi(\fft{\{\Delta_a\}_\tau}{\tau}-1)}{\psi(\fft{1-\{\Delta_a\}_\tau}{\tau}+1)}\right)\\
	&\quad+4\log(1-e^{-\fft{\pi i}{\tau}})+\mathcal O(e^{-\fft{2\pi\sin(\arg\tau)}{|\tau|}})
\end{split}\label{I:210:Cardy:kZ}
\end{equation}
and
\begin{equation}
\begin{split}
    -\log H_{\{2,1,0\}}&=-\log4-\log(\fft{\eta_2-\eta_1}{\tau}+\fft2\tau\sum_\Delta\left(\fft{e^{-\fft{2\pi i}{\tau}(1-\{\fft12+\Delta\}_\tau)}}{1-e^{-\fft{2\pi i}{\tau}(1-\{\fft12+\Delta\}_\tau)}}-\fft{e^{-\fft{2\pi i}{\tau}\{\fft12+\Delta\}_\tau}}{1-e^{-\fft{2\pi i}{\tau}\{\fft12+\Delta\}_\tau}}\right))\\
    &\quad+\mathcal O(e^{-\fft{2\pi\sin(\arg\tau)}{|\tau|}}).
\end{split}\label{I:210:Cardy:H}
\end{equation}
Here we have also used the identity (\ref{Bernoulli:identity}). The determinant contribution (\ref{I:210:Cardy:H}) now explains why we should keep track of the leading exponentially suppressed terms. If $\eta_1=\eta_2$, we get a divergent logarithmic contribution ``$\log 0$'' without those terms. The Cardy-like limit of the contribution (\ref{eq:SCI:BA:N=2:210:s=t}) is then given as the sum of (\ref{I:210:Cardy:kZ}) and (\ref{I:210:Cardy:H}), which results in (\ref{I:210:Cardy}).

Similarly the building blocks of (\ref{eq:SCI:BA:N=2:121:s=t}), or equivalently those of (\ref{eq:SCI:BA:N=2:121}) with $a=b=1$, are given as
\begin{equation}
\begin{split}
    \log\kappa\mathcal Z_{\{1,2,1\}}&=-\fft{\pi i}{2\tau^2}\prod_{a=1}^3\left(\{2\Delta_a\}_\tau-\fft{1+\eta_2}{2}\right)+\fft{\pi i}{\tau^2}\prod_{a=1}^3\left(\{\Delta_a\}_\tau-\fft{1+\eta_1}{2}\right)\\
	&\quad+\fft{\pi i(6+8\eta_1-13\eta_2)}{12}-\log\tau-\log 2!\\
	&\quad+\sum_{a=1}^3\left(\log\fft{\psi(\fft{\{1/2+\Delta_a\}_\tau}{\tau}-\fft12)}{\psi(\fft{1-\{1/2+\Delta_a\}_\tau}{\tau}+\fft12)}+\log\fft{\psi(\fft{\{1/2+\Delta_a\}_\tau}{\tau}-\fft32)}{\psi(\fft{1-\{1/2+\Delta_a\}_\tau}{\tau}+\fft32)}+\log\fft{\psi(\fft{\{\Delta_a\}_\tau}{\tau}-1)}{\psi(\fft{1-\{\Delta_a\}_\tau}{\tau}+1)}\right)\\
	&\quad+4\log(1+e^{\fft{\pi i}{\tau}})+\mathcal O(e^{-\fft{2\pi\sin(\arg\tau)}{|\tau|}})
\end{split}\label{I:121:Cardy:kZ}
\end{equation}
and
\begin{equation}
\begin{split}
    -\log H_{\{1,2,1\}}&=-\log4-\log(\fft{\eta_2-\eta_1}{\tau}+\fft2\tau\sum_\Delta\left(\fft{-e^{-\fft{2\pi i}{\tau}(1-\{\fft12+\Delta\}_\tau)}}{1+e^{-\fft{2\pi i}{\tau}(1-\{\fft12+\Delta\}_\tau)}}-\fft{-e^{-\fft{2\pi i}{\tau}\{\fft12+\Delta\}_\tau}}{1+e^{-\fft{2\pi i}{\tau}\{\fft12+\Delta\}_\tau}}\right))\\
    &\quad+\mathcal O(e^{-\fft{2\pi\sin(\arg\tau)}{|\tau|}}).
\end{split}\label{I:121:Cardy:H}
\end{equation}
The Cardy-like limit of the contribution (\ref{eq:SCI:BA:N=2:121:s=t}) is then given as the sum of (\ref{I:121:Cardy:kZ}) and (\ref{I:121:Cardy:H}), which leads to (\ref{I:121:Cardy}).

Sum of the two contributions (\ref{I:210:Cardy}) and (\ref{I:121:Cardy}) also involves complicated calculations. The resulting sum (\ref{I:Cardy:W:N=2:X}) is therefore written in terms of $X^\text{SU(2)}$, which is defined as
\begin{equation}
\begin{split}
    X^\text{SU(2)}&=\fft{(1-e^{-\fft{\pi i}{\tau}})^4\prod_{a=1}^3\fft{\psi(\fft{\{1/2+\Delta_a\}_\tau}{\tau}-1)^2}{\psi(\fft{1-\{1/2+\Delta_a\}_\tau}{\tau}+1)^2}}{\sum_\Delta\Big(\fft{e^{-\fft{2\pi i}{\tau}(1-\{\fft12+\Delta\}_\tau)}}{1-e^{-\fft{2\pi i}{\tau}(1-\{\fft12+\Delta\}_\tau)}}-\fft{e^{-\fft{2\pi i}{\tau}\{\fft12+\Delta\}_\tau}}{1-e^{-\fft{2\pi i}{\tau}\{\fft12+\Delta\}_\tau}}\Big)}\\
	&\quad+\fft{(1+e^{-\fft{\pi i}{\tau}})^4\prod_{a=1}^3\fft{\psi(\fft{\{1/2+\Delta_a\}_\tau}{\tau}-\fft12)}{\psi(\fft{1-\{1/2+\Delta_a\}_\tau}{\tau}+\fft12)}\fft{\psi(\fft{\{1/2+\Delta_a\}_\tau}{\tau}-\fft32)}{\psi(\fft{1-\{1/2+\Delta_a\}_\tau}{\tau}+\fft32)}}{\sum_\Delta\Big(\fft{-e^{-\fft{2\pi i}{\tau}(1-\{\fft12+\Delta\}_\tau)}}{1+e^{-\fft{2\pi i}{\tau}(1-\{\fft12+\Delta\}_\tau)}}-\fft{-e^{-\fft{2\pi i}{\tau}\{\fft12+\Delta\}_\tau}}{1+e^{-\fft{2\pi i}{\tau}\{\fft12+\Delta\}_\tau}}\Big)}.
\end{split}\label{eq:X:SU(2)}
\end{equation}
To simplify this expression in the Cardy-like limit, first note that
\begin{equation}
\begin{split}
    \max_\Delta\left\{|e^{-\fft{2\pi i}{\tau}(1-\{\fft12+\Delta\}_\tau)}|,|e^{-\fft{2\pi i}{\tau}\{\fft12+\Delta\}_\tau}|\right\}&=|e^{-\fft{2\pi i}{\tau}\Delta^\text{SU(2)}}|\\
    &=\begin{cases}
    |e^{-\fft{2\pi i}{\tau}\{\fft12+\Delta_3\}_\tau}| & (\eta_1=\eta_2=-1)\\
    |e^{-\fft{2\pi i}{\tau}(1-\{\fft12+\Delta_1\}_\tau)}| & (\eta_1=\eta_2=1)
    \end{cases},
\end{split}\label{Delta:SU(2):origin}
\end{equation}
under the ordering (without loss of generality)
\begin{equation}
    0<\{\tilde\Delta_1\}<\{\tilde\Delta_2\}<\{\tilde\Delta_3\}<1.
\end{equation}
This explains the origin of the definition of $\Delta^\text{SU(2)}$ given in (\ref{Delta:SU(2)}). For this $\Delta^\text{SU(2)}$, one can prove the following asymptotic expansions
\begin{equation}
\begin{split}
    \fft{\psi(\fft{\Delta^\text{SU(2)}}{\tau}-1)^{2}}{-\fft{e^{-\fft{2\pi i}{\tau}\Delta^\text{SU(2)})}}{1-e^{-\fft{2\pi i}{\tau}\Delta^\text{SU(2)})}}}+\fft{\psi(\fft{\Delta^\text{SU(2)}}{\tau}-\fft12)\psi(\fft{\Delta^\text{SU(2)}}{\tau}-\fft32)}{-\fft{-e^{-\fft{2\pi i}{\tau}\Delta^\text{SU(2)})}}{1+e^{-\fft{2\pi i}{\tau}\Delta^\text{SU(2)})}}}&\sim\fft{4\Delta^\text{SU(2)}}{\tau}-2-\fft{2i}{\pi},\\
    \fft{\psi(\fft{\Delta^\text{SU(2)}}{\tau}+1)^{-2}}{\fft{e^{-\fft{2\pi i}{\tau}\Delta^\text{SU(2)})}}{1-e^{-\fft{2\pi i}{\tau}\Delta^\text{SU(2)})}}}+\fft{\psi(\fft{\Delta^\text{SU(2)}}{\tau}+\fft12)^{-1}\psi(\fft{\Delta^\text{SU(2)}}{\tau}+\fft32)^{-1}}{\fft{-e^{-\fft{2\pi i}{\tau}\Delta^\text{SU(2)})}}{1+e^{-\fft{2\pi i}{\tau}\Delta^\text{SU(2)})}}}&\sim\fft{4\Delta^\text{SU(2)}}{\tau}+2-\fft{2i}{\pi},
\end{split}\label{eq:SU(2):tool}
\end{equation}
using the expansion of the $\psi$-function (\ref{psi:asymp}). Applying the above results (\ref{eq:SU(2):tool}) to the definition of $X^\text{SU(2)}$ (\ref{eq:X:SU(2)}) along with (\ref{Delta:SU(2):origin}) and (\ref{Delta:SU(2)}), we obtain the asymptotic expansion of $X^\text{SU(2)}$ (\ref{eq:X:SU(2):approx}).

\subsection{SU(3) case}\label{App:Cardy:N=3}
For the SU(3) case, there are three remaining standard BAE solutions: $\{3,1,0\}$ and $\{1,3,1\}$, and $\{1,3,2\}$. Their contributions to the SCI are given in (\ref{eq:SCI:BA:N=3:310}), (\ref{eq:SCI:BA:N=3:131}), and (\ref{eq:SCI:BA:N=3:132}) with $p=q~(a=b=1)$ respectively.

First, the building blocks of (\ref{eq:SCI:BA:N=3:310}) with $a=b=1$ can be computed using the asymptotic expansions (\ref{elliptic:theta:1:asymp}) and (\ref{elliptic:Gamma:asymp}) as
\begin{equation}
\begin{split}
	&\log\kappa\mathcal Z_{\{3,1,0\}}\\
	&=-\fft{\pi i}{3\tau^2}\prod_{a=1}^3\left(\{3\Delta_a\}_\tau-\fft{1+\eta_3}{2}\right)+\fft{\pi i}{\tau^2}\prod_{a=1}^3\left(\{\Delta_a\}_\tau-\fft{1+\eta_1}{2}\right)\\
	&\quad+\fft{\pi i(12+5\eta_1-15\eta_3)}{12}-2\log\tau-\log 3!\\
	&\quad+\sum_{a=1}^3\left(3\log\fft{\psi(\fft{\{1/3+\Delta_a\}_\tau}{\tau}-1)}{\psi(\fft{1-\{1/3+\Delta_a\}_\tau}{\tau}+1)}+3\log\fft{\psi(\fft{\{2/3+\Delta_a\}_\tau}{\tau}-1)}{\psi(\fft{1-\{2/3+\Delta_a\}_\tau}{\tau}+1)}+2\log\fft{\psi(\fft{\{\Delta_a\}_\tau}{\tau}-1)}{\psi(\fft{1-\{\Delta_a\}_\tau}{\tau}+1)}\right)\\
	&\quad+6\log(1-e^{-\fft{2\pi i}{3\tau}})+6\log(1-e^{-\fft{4\pi i}{3\tau}})+\mathcal O(e^{-\fft{2\pi\sin(\arg\tau)}{|\tau|}})
\end{split}\label{I:310:Cardy:kZ}%
\end{equation}
and
\begin{equation}
\begin{split}
	&-\log H_{\{3,1,0\}}\\
	&=-3\log3-2\log(\fft{\eta_3-\eta_1}{2\tau}+\fft1\tau\sum_{J=1}^2\sum_\Delta\left(\fft{e^{-\fft{2\pi i}{\tau}(1-\{\fft{J}{3}+\Delta\}_\tau)}}{1-e^{-\fft{2\pi i}{\tau}(1-\{\fft{J}{3}+\Delta\}_\tau)}}-\fft{e^{-\fft{2\pi i}{\tau}\{\fft{J}{3}+\Delta\}_\tau}}{1-e^{-\fft{2\pi i}{\tau}\{\fft{J}{3}+\Delta\}_\tau}}\right)).\label{I:310:Cardy:H}
\end{split}
\end{equation}
Here we have also used the identity (\ref{Bernoulli:identity}). The determinant contribution (\ref{I:310:Cardy:H}) now explains why we should keep track of the leading exponentially suppressed terms. If $\eta_1=\eta_3$, we get a logarithmically divergent contribution ``$\log 0$'' without those terms. The Cardy-like limit of the contribution (\ref{eq:SCI:BA:N=3:310}) with $a=b=1$ is then given as the sum of (\ref{I:310:Cardy:kZ}) and (\ref{I:310:Cardy:H}), which gives (\ref{I:310:Cardy}).

Similarly the building blocks of (\ref{eq:SCI:BA:N=3:131}) with $a=b=1$ are given as
\begin{equation}
\begin{split}
	&\log\kappa\mathcal Z_{\{1,3,1\}}\\
	&=-\fft{\pi i}{3\tau^2}\prod_{a=1}^3\left(\{3\Delta_a\}_\tau-\fft{1+\eta_3}{2}\right)+\fft{\pi i}{\tau^2}\prod_{a=1}^3\left(\{\Delta_a\}_\tau-\fft{1+\eta_1}{2}\right)\\
	&\quad+\fft{\pi i(12+9\eta_1-19\eta_3)}{12}-2\log\tau-\log 3!\\
	&\quad+\sum_{a=1}^3\left(2\log\fft{\psi(\fft{\{1/3+\Delta_a\}_\tau}{\tau}-\fft23)}{\psi(\fft{1-\{1/3+\Delta_a\}_\tau}{\tau}+\fft23)}+2\log\fft{\psi(\fft{\{2/3+\Delta_a\}_\tau}{\tau}-\fft43)}{\psi(\fft{1-\{2/3+\Delta_a\}_\tau}{\tau}+\fft43)}\right.\\
	&\kern4em~\left.+\log\fft{\psi(\fft{\{2/3+\Delta_a\}_\tau}{\tau}-\fft13)}{\psi(\fft{1-\{2/3+\Delta_a\}_\tau}{\tau}+\fft13)}+\log\fft{\psi(\fft{\{1/3+\Delta_a\}_\tau}{\tau}-\fft53)}{\psi(\fft{1-\{1/3+\Delta_a\}_\tau}{\tau}+\fft53)}+2\log\fft{\psi(\fft{\{\Delta_a\}_\tau}{\tau}-1)}{\psi(\fft{1-\{\Delta_a\}_\tau}{\tau}+1)}\right)\\
	&\quad+6\log(1-e^{-2\pi i(\fft{1}{3\tau}-\fft23)})+6\log(1-e^{-2\pi i(\fft{2}{3\tau}-\fft13)})+\mathcal O(e^{-\fft{2\pi\sin(\arg\tau)}{|\tau|}})
\end{split}\label{I:131:Cardy:kZ}%
\end{equation}
and
\begin{equation}
\begin{split}
	&-\log H_{\{1,3,1\}}\\
	&=-3\log3-2\log(\fft{\eta_3-\eta_1}{2\tau}+\fft1\tau\sum_{J=1}^2\sum_\Delta\left(\fft{e^{-\fft{2\pi i}{\tau}(1-\{\fft{J}{3}+\Delta\}_\tau-\fft{J\tau}{3})}}{1-e^{-\fft{2\pi i}{\tau}(1-\{\fft{J}{3}+\Delta\}_\tau-\fft{J\tau}{3})}}-\fft{e^{-\fft{2\pi i}{\tau}(\{\fft{J}{3}+\Delta\}_\tau+\fft{J\tau}{3})}}{1-e^{-\fft{2\pi i}{\tau}(\{\fft{J}{3}+\Delta\}_\tau+\fft{J\tau}{3})}}\right)).\label{I:131:Cardy:H}
\end{split}
\end{equation}
The Cardy-like limit of the contribution (\ref{eq:SCI:BA:N=3:131}) with $a=b=1$ is then given as the sum of (\ref{I:131:Cardy:kZ}) and (\ref{I:131:Cardy:H}), which gives (\ref{I:131:Cardy}).

Finally the building blocks of (\ref{eq:SCI:BA:N=3:132}) with $a=b=1$ are given as
\begin{equation}
\begin{split}
	&\log\kappa\mathcal Z_{\{1,3,2\}}\\
	&=-\fft{\pi i}{3\tau^2}\prod_{a=1}^3\left(\{3\Delta_a\}_\tau-\fft{1+\eta_3}{2}\right)+\fft{\pi i}{\tau^2}\prod_{a=1}^3\left(\{\Delta_a\}_\tau-\fft{1+\eta_1}{2}\right)\\
	&\quad+\fft{\pi i(12+9\eta_1-19\eta_3)}{12}-2\log\tau-\log 3!\\
	&\quad+\sum_{a=1}^3\left(2\log\fft{\psi(\fft{\{2/3+\Delta_a\}_\tau}{\tau}-\fft23)}{\psi(\fft{1-\{2/3+\Delta_a\}_\tau}{\tau}+\fft23)}+2\log\fft{\psi(\fft{\{1/3+\Delta_a\}_\tau}{\tau}-\fft43)}{\psi(\fft{1-\{1/3+\Delta_a\}_\tau}{\tau}+\fft43)}\right.\\
	&\kern4em~\left.+\log\fft{\psi(\fft{\{1/3+\Delta_a\}_\tau}{\tau}-\fft13)}{\psi(\fft{1-\{1/3+\Delta_a\}_\tau}{\tau}+\fft13)}+\log\fft{\psi(\fft{\{2/3+\Delta_a\}_\tau}{\tau}-\fft53)}{\psi(\fft{1-\{2/3+\Delta_a\}_\tau}{\tau}+\fft53)}+2\log\fft{\psi(\fft{\{\Delta_a\}_\tau}{\tau}-1)}{\psi(\fft{1-\{\Delta_a\}_\tau}{\tau}+1)}\right)\\
	&\quad+6\log(1-e^{-2\pi i(\fft{2}{3\tau}-\fft23)})+6\log(1-e^{-2\pi i(\fft{1}{3\tau}-\fft13)})+\mathcal O(e^{-\fft{2\pi\sin(\arg\tau)}{|\tau|}}),
\end{split}\label{I:132:Cardy:kZ}%
\end{equation}
and
\begin{equation}
\begin{split}
	&-\log H_{\{1,3,2\}}\\
	&=-3\log3-2\log(\fft{\eta_3-\eta_1}{2\tau}+\fft1\tau\sum_{J=1}^2\sum_\Delta\left(\fft{e^{-\fft{2\pi i}{\tau}(1-\{\fft{J}{3}+\Delta\}_\tau+\fft{J\tau}{3})}}{1-e^{-\fft{2\pi i}{\tau}(1-\{\fft{J}{3}+\Delta\}_\tau+\fft{J\tau}{3})}}-\fft{e^{-\fft{2\pi i}{\tau}(\{\fft{J}{3}+\Delta\}_\tau-\fft{J\tau}{3})}}{1-e^{-\fft{2\pi i}{\tau}(\{\fft{J}{3}+\Delta\}_\tau-\fft{J\tau}{3})}}\right)).\label{I:132:Cardy:H}
\end{split}
\end{equation}
The Cardy-like limit of the contribution (\ref{eq:SCI:BA:N=3:132}) with $a=b=1$ is then given as the sum of (\ref{I:132:Cardy:kZ}) and (\ref{I:132:Cardy:H}), which gives (\ref{I:132:Cardy}).

Sum of the three contributions (\ref{I:310:Cardy}), (\ref{I:131:Cardy}), and (\ref{I:132:Cardy}) also involves complicated calculations. The resulting sum (\ref{I:Cardy:W:N=3:X}) is therefore written in terms of $X^\text{SU(3)}$, which is defined as
\begin{equation}
\begin{split}
	&X^\text{SU(3)}\\
	&=\fft{(1-e^{-\fft{2\pi i}{3\tau}})^6(1-e^{-\fft{4\pi i}{3\tau}})^6\prod_{a=1}^3\fft{\psi(\fft{\{1/3+\Delta_a\}_\tau}{\tau}-1)^3}{\psi(\fft{1-\{1/3+\Delta_a\}_\tau}{\tau}+1)^3}\fft{\psi(\fft{\{2/3+\Delta_a\}_\tau}{\tau}-1)^3}{\psi(\fft{1-\{2/3+\Delta_a\}_\tau}{\tau}+1)^3}}{\left(\sum_{J=1}^2\sum_\Delta(\fft{e^{-\fft{2\pi i}{\tau}(1-\{J/3+\Delta\}_\tau)}}{1-e^{-\fft{2\pi i}{\tau}(1-\{J/3+\Delta\}_\tau)}}-\fft{e^{-\fft{2\pi i}{\tau}\{J/3+\Delta\}_\tau}}{1-e^{-\fft{2\pi i}{\tau}\{J/3+\Delta\}_\tau}})\right)^2}\\
	&\quad+\fft{(1-e^{-2\pi i(\fft{1}{3\tau}-\fft23)})^6(1-e^{-2\pi i(\fft{2}{3\tau}-\fft13)})^6}{\left(\sum_{J=1}^2(\fft{e^{-\fft{2\pi i}{\tau}(1-\{J/3+\Delta\}_\tau-J\tau/3)}}{1-e^{-\fft{2\pi i}{\tau}(1-\{J/3+\Delta\}_\tau-J\tau/3)}}-\fft{e^{-\fft{2\pi i}{\tau}(\{J/3+\Delta\}_\tau+J\tau/3)}}{1-e^{-\fft{2\pi i}{\tau}(\{J/3+\Delta\}_\tau+J\tau/3)}})\right)^2}\\
	&\qquad\times\prod_{a=1}^3\fft{\psi(\fft{\{1/3+\Delta_a\}_\tau}{\tau}-\fft23)^2}{\psi(\fft{1-\{1/3+\Delta_a\}_\tau}{\tau}+\fft23)^2}\fft{\psi(\fft{\{2/3+\Delta_a\}_\tau}{\tau}-\fft43)^2}{\psi(\fft{1-\{2/3+\Delta_a\}_\tau}{\tau}+\fft43)^2}\fft{\psi(\fft{\{2/3+\Delta_a\}_\tau}{\tau}-\fft13)}{\psi(\fft{1-\{2/3+\Delta_a\}_\tau}{\tau}+\fft13)}\fft{\psi(\fft{\{1/3+\Delta_a\}_\tau}{\tau}-\fft53)}{\psi(\fft{1-\{1/3+\Delta_a\}_\tau}{\tau}+\fft53)}\\
	&\quad+\fft{(1-e^{-2\pi i(\fft{2}{3\tau}-\fft23)})^6(1-e^{-2\pi i(\fft{1}{3\tau}-\fft13)})^6}{\left(\sum_{J=1}^2(\fft{e^{-\fft{2\pi i}{\tau}(1-\{J/3+\Delta\}_\tau+J\tau/3)}}{1-e^{-\fft{2\pi i}{\tau}(1-\{J/3+\Delta\}_\tau+J\tau/3)}}-\fft{e^{-\fft{2\pi i}{\tau}(\{J/3+\Delta\}_\tau-J\tau/3)}}{1-e^{-\fft{2\pi i}{\tau}(\{J/3+\Delta\}_\tau-J\tau/3)}})\right)^2}\\
	&\qquad\times\prod_{a=1}^3\fft{\psi(\fft{\{2/3+\Delta_a\}_\tau}{\tau}-\fft23)^2}{\psi(\fft{1-\{2/3+\Delta_a\}_\tau}{\tau}+\fft23)^2}\fft{\psi(\fft{\{1/3+\Delta_a\}_\tau}{\tau}-\fft43)^2}{\psi(\fft{1-\{1/3+\Delta_a\}_\tau}{\tau}+\fft43)^2}\fft{\psi(\fft{\{1/3+\Delta_a\}_\tau}{\tau}-\fft13)}{\psi(\fft{1-\{1/3+\Delta_a\}_\tau}{\tau}+\fft13)}\fft{\psi(\fft{\{2/3+\Delta_a\}_\tau}{\tau}-\fft53)}{\psi(\fft{1-\{2/3+\Delta_a\}_\tau}{\tau}+\fft53)}.
\end{split}\label{eq:X:SU(3)}%
\end{equation}
To simplify this expression in the Cardy-like limit, first note that
\begin{equation}
\begin{split}
    &\max_\Delta\left\{|e^{-\fft{2\pi i}{\tau}(1-\{\fft13+\Delta\}_\tau)}|,|e^{-\fft{2\pi i}{\tau}(1-\{\fft23+\Delta\}_\tau)}|,|e^{-\fft{2\pi i}{\tau}\{\fft13+\Delta\}_\tau}|,|e^{-\fft{2\pi i}{\tau}\{\fft23+\Delta\}_\tau}|\right\}\\
    &=|e^{-\fft{2\pi i}{\tau}\Delta^\text{SU(3)}}|=\begin{cases}
    |e^{-\fft{2\pi i}{\tau}\{\fft13+\Delta_3\}_\tau}| & (\eta_1=\eta_3=-1,~\{\tilde\Delta_3\}>2/3)\\
    |e^{-\fft{2\pi i}{\tau}\{\fft23+\Delta_2\}_\tau}| & (\eta_1=\eta_3=-1,~\{\tilde\Delta_3\}<2/3)\\
    |e^{-\fft{2\pi i}{\tau}(1-\{\fft23+\Delta_1\}_\tau)}| & (\eta_1=\eta_3=1,~\{\tilde\Delta_1\}<1/3)\\
    |e^{-\fft{2\pi i}{\tau}(1-\{\fft13+\Delta_2\}_\tau)}| & (\eta_1=\eta_3=1,~\{\tilde\Delta_1\}>1/3)
    \end{cases},
\end{split}\label{Delta:SU(3):origin}
\end{equation}
under the ordering (without loss of generality)
\begin{equation}
    0<\{\tilde\Delta_1\}<\{\tilde\Delta_2\}<\{\tilde\Delta_3\}<1.
\end{equation}
This explains the origin of the definition of $\Delta^\text{SU(3)}$ given in (\ref{Delta:SU(3)}). For this $\Delta^\text{SU(3)}$, one can prove the following asymptotic expansions
\begin{equation}
\begin{split}
    &\fft{\psi(\fft{\Delta^\text{SU(3)}}{\tau}-1)^3}{\Big(-\fft{e^{-\fft{2\pi i}{\tau}\Delta^\text{SU(3)})}}{1-e^{-\fft{2\pi i}{\tau}\Delta^\text{SU(3)})}}\Big)^2}+\fft{\psi(\fft{\Delta^\text{SU(3)}}{\tau}-\fft23)^2\psi(\fft{\Delta^\text{SU(3)}}{\tau}-\fft53)}{\Big(-\fft{w^2e^{-\fft{2\pi i}{\tau}\Delta^\text{SU(3)})}}{1-w^2e^{-\fft{2\pi i}{\tau}\Delta^\text{SU(3)})}}\Big)^2}+\fft{\psi(\fft{\Delta^\text{SU(3)}}{\tau}-\fft43)^2\psi(\fft{\Delta^\text{SU(3)}}{\tau}-\fft13)}{\Big(-\fft{we^{-\fft{2\pi i}{\tau}\Delta^\text{SU(3)})}}{1-we^{-\fft{2\pi i}{\tau}\Delta^\text{SU(3)})}}\Big)^2}\\
    &\sim\fft{27(\Delta^\text{SU(3)})^2}{2\tau^2}+\fft{27\Delta^\text{SU(3)}(-\pi-i)}{2\pi\tau}+\fft{3(8\pi^2+15\pi i-9)}{8\pi^2},\\
    &\fft{\psi(\fft{\Delta^\text{SU(3)}}{\tau}+1)^{-3}}{\Big(\fft{e^{-\fft{2\pi i}{\tau}\Delta^\text{SU(3)})}}{1-e^{-\fft{2\pi i}{\tau}\Delta^\text{SU(3)})}}\Big)^2}+\fft{\psi(\fft{\Delta^\text{SU(3)}}{\tau}+\fft23)^{-2}\psi(\fft{\Delta^\text{SU(3)}}{\tau}+\fft53)^{-1}}{\Big(\fft{we^{-\fft{2\pi i}{\tau}\Delta^\text{SU(3)})}}{1-we^{-\fft{2\pi i}{\tau}\Delta^\text{SU(3)})}}\Big)^2}+\fft{\psi(\fft{\Delta^\text{SU(3)}}{\tau}+\fft43)^{-2}\psi(\fft{\Delta^\text{SU(3)}}{\tau}+\fft13)^{-1}}{\Big(\fft{w^2e^{-\fft{2\pi i}{\tau}\Delta^\text{SU(3)})}}{1-w^2e^{-\fft{2\pi i}{\tau}\Delta^\text{SU(3)})}}\Big)^2}\\
    &\sim\fft{27(\Delta^\text{SU(3)})^2}{2\tau^2}+\fft{27\Delta^\text{SU(3)}(\pi-i)}{2\pi\tau}+\fft{3(8\pi^2-15\pi i-9)}{8\pi^2},
\end{split}\label{eq:SU(3):tool}
\end{equation}
using the expansion of the $\psi$-function (\ref{psi:asymp}). Here $w=e^{2\pi i/3}$ is a primitive cube root of unity. Applying the above results (\ref{eq:SU(3):tool}) to the definition of $X^\text{SU(3)}$ (\ref{eq:X:SU(3)}) along with (\ref{Delta:SU(3):origin}) and (\ref{Delta:SU(3)}), we obtain the asymptotic expansion of $X^\text{SU(3)}$ (\ref{eq:X:SU(3):approx}).

\bibliographystyle{JHEP}
\bibliography{BHLocalization}
\end{document}